\documentclass[aps,pra,preprintnumbers,showpacs,tightenlines]{revtex4}
\usepackage{amssymb}
\usepackage{amsmath}
\usepackage{graphicx}
\usepackage{epsfig}
\usepackage{subfigure}
\usepackage{amsfonts}
\usepackage{CJK}

\begin{document}

\title{Phase gate of one qubit simultaneously controlling $n$ qubits in a cavity}

\author{Chui-Ping Yang$^{1,2}$, Yu-xi Liu$^{3,4,1}$, and Franco Nori$^{1,2}$}

\address{$^1$Advanced Science Institute, The Institute of
Physical and Chemical Research (RIKEN), Wako-Shi, Saitama
351-0198, Japan}

\address{$^2$Physics Department, The University of Michigan,
Ann Arbor, Michigan 48109-1040, USA}

\address{$^3$Institute of Microelectronics, Tsinghua
University, Beijing 100084, China}

\address{$^4$Tsinghua National Laboratory for Information
Science and Technology (TNList), Tsinghua University, Beijing
100084, China}

\date{\today}

\begin{abstract}
We propose how to realize a three-step controlled-phase gate of
one qubit simultaneously controlling $n$ qubits in a cavity or
coupled to a resonator. The $n$ two-qubit controlled-phase gates,
forming this multiqubit phase gate, can be performed
simultaneously. The operation time of this phase gate is
independent of the number $n$ of qubits. This phase gate
controlling at once $n$ qubits is insensitive to the initial state
of the cavity mode and can be used to produce an analogous CNOT
gate simultaneously acting on $n$ qubits. We present two
alternative approaches to implement this gate. One approach is
based on tuning the qubit frequency while the other method tunes
the resonator frequency. Using superconducting qubits coupled to a
resonator as an example, we show how to implement the proposed
gate with one superconducting qubit simultaneously controlling $n$
qubits selected from $N$ qubits coupled to a resonator ($1<n<N$).
We also give a discussion on realizing the proposed gate with
atoms, by using one cavity initially in an arbitrary state.
\end{abstract}

 \maketitle
\date{\today}

\begin{center}
\textbf{I. INTRODUCTION}
\end{center}

Quantum information processing has attracted considerable interest during
the past decade. The building blocks of quantum computing are single-qubit
and two-qubit logic gates. So far, a large number of theoretical proposals
for realizing two-qubit gates in many physical systems, such as trapped
ions, atoms in cavity QED, nuclear magnetic resonance (NMR), and solid-state
devices, have been proposed.~Moreover, two-qubit controlled-not (CNOT),
controlled-phase (CP), $i$SWAP gates, or other two-qubit operations have
been experimentally demonstrated in ion traps~[1], NMR~[2], quantum
dots~[3], and superconducting qubits~[4-8].

In recent years, analogs of cavity quantum electrodynamics (QED) have been
studied using superconducting Josephson devices. There, a superconducting
resonator provides a quantized cavity field, superconducting qubits act as
artificial atoms, and the strong coupling between the field and
superconducting qubits has been demonstrated [9]. Theoretically, many
approaches for performing two-qubit logic gates using superconducting qubits
coupled to a superconducting resonator or cavity have been proposed [e.g.,
10-15]. Moveover, experimental demonstrations of two-qubit gates or
two-qubit operations using superconducting qubits coupled to cavities have
recently been reported [7,8].

Attention is now shifting to the physical realization of \textit{multi}%
-qubit gates (see,~e.g.,~Ref.~[16]) instead of just \textit{two}-qubit
gates. It is known that any multi-qubit gate can be decomposed into
two-qubit gates and one-qubit gates. When using the conventional
gate-decomposition protocols to construct a multi-qubit gate~[17,18], the
procedure usually becomes complicated as the number of qubits increases.
Therefore, building a multi-qubit gate may become very difficult since each
elementary gate requires turning on and off a given Hamiltonian for a
certain period of time, and each additional basic gate adds experimental
complications and the possibility of more errors. During the past few years,
several methods for constructing multi-qubit phase gates with $n$-control
qubits acting on one target qubit based on ion traps [19], cavity QED
[20,21], or circuit QED [22] have been proposed. As is well known, a
multi-qubit gate with multiple control qubits controlling a single qubit
plays a significant role in quantum information processing, such as quantum
algorithms [e.g.,23,24] and quantum error-correction protocols [25].

\begin{figure}[tbp]
\includegraphics[bb=129 250 441 703, width=8.0 cm, clip]{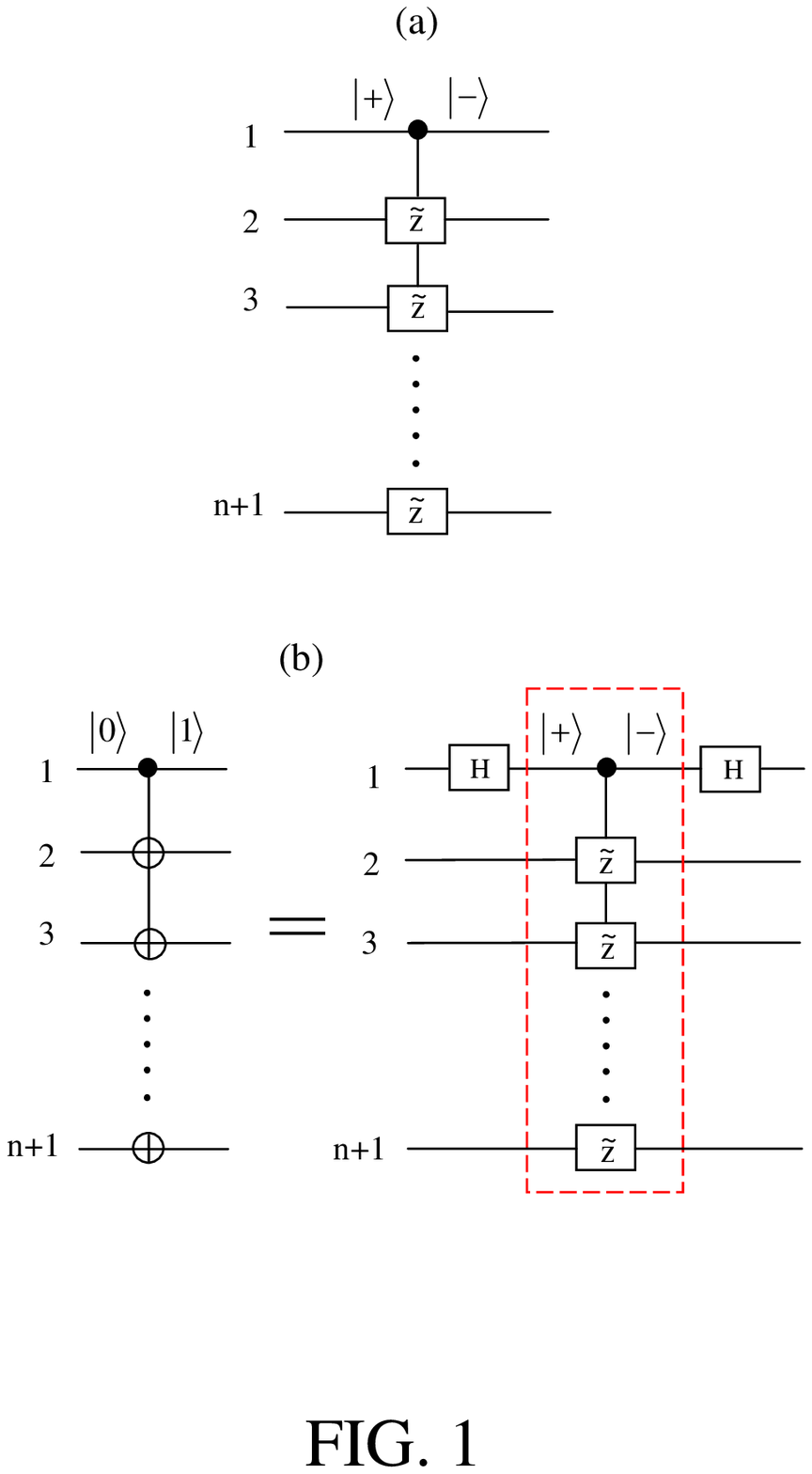} %
\vspace*{-0.08in}
\caption{(Color online) (a) Schematic circuit of a $n$-target-qubit control
phase (NTCP) gate with qubit 1 simultaneously controlling $n$ target qubits
(2,~3,~...,~$n+1$). The NTCP gate is equivalent to $n$ two-qubit control
phase (CP) gates each having a shared control qubit (qubit 1) but a
different target qubit (either qubit 2,~3,~...,~or~$n+1$). ~Here, $%
\widetilde{Z}$ represents a controlled-phase flip on each target qubit, with
respect to the computational basis formed by the two eigenstates $\left|
+\right\rangle$ and $\left| -\right\rangle$ of the Pauli operator $\sigma
_x. $~Namely, if the control qubit 1 is in the state $\left| -\right\rangle$%
, then the state $\left| -\right\rangle$ at each $\widetilde{Z}$ is
phase-flipped as $\left| -\right\rangle$ $\rightarrow -\left| -\right\rangle$%
, while the state $\left| +\right\rangle$ remains unchanged. (b)
Relationship between a $n$-target-qubit controlled-NOT gate and a NTCP gate.
The circuit on the left side of (b) is equivalent to the circuit on the
right side of (b). For the circuit on the left side, the symbol $\oplus $
represents a NOT gate on each target qubit, with respect to the
computational basis formed by the two eigenstates $\left| 0\right\rangle $
and $\left| 1\right\rangle $ of the Pauli operator $\sigma _z$.~If the
control qubit 1 is in the state $\left| 1\right\rangle $, then the state at $%
\oplus $ is bit flipped as $\left| 1\right\rangle $ $\rightarrow \left|
0\right\rangle $ and $\left| 0\right\rangle $ $\rightarrow \left|
1\right\rangle $.~However,~when the control qubit 1 is in the state $\left|
0\right\rangle $,~the state at $\oplus $ remains unchanged. On the other
hand, for the circuit on the right side, the part enclosed in the (red)
dashed-line box represents a NTCP gate. The element containing H corresponds
to a Hadamard transformation described by $\left| 0\right\rangle \rightarrow
\left| +\right\rangle=\left( 1/\protect\sqrt{2}\right) \left( \left|
0\right\rangle +\left| 1\right\rangle \right) $, and $\left| 1\right\rangle
\rightarrow \left| -\right\rangle =\left( 1/\protect\sqrt{2}\right) \left(
\left| 0\right\rangle -\left| 1\right\rangle \right) .$}
\label{fig:1}
\end{figure}

In this work, we focus on another type of multi-qubit gates, i.e., a
multi-qubit phase gate with \textit{one} control qubit simultaneously
controlling $n$ target qubits. This multi-qubit gate is useful in quantum
information processing such as entanglement preparation~[26], error
correction~[27], quantum algorithms (e.g., the Discrete Cosine
Transform~[28]), and quantum cloning~[29]. In the following, we will propose
a way for realizing this multi-qubit gate using ($n+1$) qubits in a cavity
or coupled to a resonator.~To implement this gate, we construct an effective
Hamiltonian which contains interaction terms between the control qubit and
each subordinate or target qubit. We will denote this $n$-target-qubit
control-phase gate as an NTCP gate [Fig.~1(a)]. We present two alternative
approaches to implement this NTCP gate. One approach is based on tuning the
qubit frequency, while the other method tunes the cavity (or resonator)
frequency. We present these two alternative methods because some
experimental implementations might find it easier to tune the qubit
frequency, while others might prefer to tune the cavity frequency.~For solid
state qubits such as superconducting qubits and semiconductor quantum dots,
the qubit transition frequency (or the qubit level spacings) can be readily
adjusted by varying the external parameters [30-34] (e.g., the external
magnetic flux for superconducting charge qubits, the flux bias or current
bias in the case of superconducting phase qubits and flux qubits, see e.g.
[30-33], and the external electric field for semiconductor quantum dots
[34]). Also, the cavity mode frequency can be changed in various experiments
(e.g., [35-39]).

As shown below, our proposal has the following advantages: (i) The $n$
two-qubit CP gates involved in the NTCP gate can be performed
simultaneously; (ii) The operation time required for the gate implementation
is independent of the number $n$ of qubits; (iii) This proposal is
insensitive to the initial state of the cavity mode, and thus no preparation
for the initial state of the cavity mode is needed; (iv) No measurement on
the qubits or the cavity mode is needed and thus the operation is
simplified; and (v) The proposal requires only three steps of operations.

Note that a CNOT gate of one qubit simultaneously controlling $n$ qubits,
shown in Fig.~1(b), can also be achieved using the present proposal. This is
because the $n$-target-qubit CNOT gate is equivalent to an NTCP gate plus
two Hadamard gates on the control qubit [Fig. 1(b)].

To the best of our knowledge, our proposal is the only one so far to
demonstrate that a powerful phase gate, synchronously controlling $n$
qubits, can be achieved in a cavity or resonator, which can be initially in
an arbitrary state. This proposal is quite general and can be applied to
physical systems such as trapped atoms, quantum dots, and superconducting
qubits. We believe that this work is of general interest and significance
because it provides a protocol for performing a controlled-phase (or
controlled-not) gate with multiple target qubits.

This paper is organized as follows. In Sec.~II, we introduce the $n$%
-target-qubit control phase gate studied in this work. In Sec.~III, we
discuss how to obtain the time-evolution operators for a qubit system
interacting with a single cavity mode and driven by a classical pulse. In
Sec.~IV, using the time-evolution operators obtained, we present two
alternative approaches for realizing the NTCP gate with qubits in a cavity
or coupled to a resonator. In the appendix, we provide guidelines on how to
protect multi-level qubits from leaking out of the computational subspace.
In Sec.~V, using superconducting qubits coupled to a resonator, we show how
to apply our general proposal to implement the proposed NTCP gate, and then
discuss its feasibility based on current experiments in superconducting
quantum circuits. In Sec. VI, we discuss how to extend the present proposal
to implement the NTCP gate with trapped atomic qubits, by using one cavity.
A concluding summary is provided in Sec.~VII.

\begin{center}
\textbf{II. ONE QUBIT SIMULTANEOUSLY CONTROLLING {\Large $n$} TARGET QUBITS}
\end{center}

For two qubits, there are a total of four computational basis states,
denoted by $\left| ++\right\rangle ,$ $\left| +-\right\rangle ,$ $\left|
-+\right\rangle ,$ and $\left| --\right\rangle .$ Here, $\left|
+\right\rangle $ and $\left| -\right\rangle $ are the two eigenstates of the
Pauli operator $\sigma _x$. A two-qubit CP gate is defined as follows
\begin{eqnarray}
\left| ++\right\rangle  &\rightarrow &\left| ++\right\rangle ,\;\text{ }%
\left| +-\right\rangle \rightarrow \left| +-\right\rangle ,\text{ }
\nonumber \\
\left| -+\right\rangle  &\rightarrow &\left| -+\right\rangle ,\;\text{ }%
\left| --\right\rangle \rightarrow -\left| --\right\rangle ,
\end{eqnarray}
which implies that if and only if the control qubit (the first qubit) is in
the state $\left| -\right\rangle $, a phase flip happens to the state $%
\left| -\right\rangle $ of the target qubit (the second qubit), but nothing
happens otherwise.

The NTCP gate considered here consists of $n$ two-qubit CP gates [Fig.
1(a)]. Each two-qubit CP gate involved in this NTCP gate has a \textit{shared%
} control qubit (labelled by $1$) but a different target qubit (labelled by $%
2,3,...,$ or $n+1$). According to the transformation (1) for a two-qubit CP
gate, it can be seen that this NTCP gate with one control qubit (qubit $1$)
and $n$ target qubits (qubits $2,3,...,n+1$) can be described by the
following unitary operator

\begin{equation}
U=\prod_{j=2}^{n+1}\left( I_j-2\left| -_1-_j\right\rangle \left\langle
-_1-_j\right| \right) ,
\end{equation}
where the subscript $1$ represents the control qubit $1$, while $j$
represents the $j$th target qubit, and $I_j$ is the identity operator for
the qubit pair ($1,j$), which is given by $I_j=\sum_{\mathrm{rs}}\left|
\mathrm{r}_1\mathrm{s}_j\right\rangle \left\langle \mathrm{r}_1\mathrm{s}%
_j\right| $, with \textrm{r, s }$\in \left\{ +,-\right\} $. One can see that
the operator (2) induces a phase flip (from the $+$ sign to the $-$ sign) to
the logical state $\left| -\right\rangle $ of each target qubit when the
control qubit $1$ is initially in the state $\left| -\right\rangle $, and
nothing happens otherwise.

It should be mentioned that the NTCP gate can be defined using the
eigenstates $\left| 0\right\rangle $ and $\left| 1\right\rangle $ of the
Pauli operator $\sigma _z.$ However, we note that to construct a $n$-target
controlled-NOT gate (based on the NTCP gate defined in the eigenstates of $%
\sigma _z$), a Hadamard gate acting on each target qubit before and after
the NTCP gate (i.e., a total of $2n$ Hadamard gates) would be required. In
contrast, the construction of the $n$-target controlled-NOT gate, using the
NTCP gate defined in the eigenstates of $\sigma _x,$ requires only $2$
Hadamard gates [see Fig. 1(b)]. Therefore, the NTCP here, defined in the
eigenstates of $\sigma _x$, makes the procedure for constructing a $n$%
-target controlled-NOT gate much simpler.

\begin{center}
\textbf{III. MODEL AND UNITARY EVOLUTION}
\end{center}

Consider ($n+1$) qubits interacting with the cavity mode and driven by a
classical pulse. In the rotating-wave approximation, the Hamiltonian for the
whole system is
\begin{equation}
H=H_0+H_1+H_2,
\end{equation}
with
\begin{eqnarray}
H_0 &=&-\frac{\hbar \omega _0}2S_z+\hbar \omega _ca^{\dagger }a, \\
H_1 &=&\frac{\hbar \Omega }2\left[ e^{i\left( \omega t+\varphi \right)
}S_{-}+e^{-i\left( \omega t+\varphi \right) }S_{+}\right] , \\
H_2 &=&\hbar g\left( aS_{+}+a^{\dagger }S_{-}\right) .
\end{eqnarray}
The Hamiltonian (3), together with the Hamiltonians (4-6), can be
implemented when the qubits are atoms [40,41], quantum dots or
superconducting devices (e.g., see section V below). Here, $H_0$ is the free
Hamiltonian of the qubits and the cavity mode, $H_1$ is the interaction
Hamiltonian between the qubits and the classical pulse, and $H_2$ is the
interaction Hamiltonian between the qubits and the cavity mode. In addition,
$\omega _0$ is the transition frequency between the two levels $\left|
0\right\rangle $ and $\left| 1\right\rangle $ of each qubit; $a$ ($%
a^{\dagger }$) is the photon annihilation (creation) operator of the cavity
mode with frequency $\omega _c;$ $g$ is the coupling constant between the
cavity mode and each qubit; $\Omega ,$ $\omega ,$ and $\varphi $ are the
Rabi frequency, the frequency, and the initial phase of the pulse,
respectively; and $S_z,$ $S_{-},$ and $S_{+}$ are the collective operators
for the $(n+1)$ qubits, which are given by
\begin{equation}
S_z=\sum_{j=1}^{n+1}\sigma _{z,j},\text{\ }S_{-}=\sum_{j=1}^{n+1}\sigma
_j^{-},\;S_{+}=\sum_{j=1}^{n+1}\sigma _j^{+},
\end{equation}
where $\sigma _{z,j}=\left| 0_j\right\rangle \left\langle 0_j\right| -\left|
1_j\right\rangle \left\langle 1_j\right| ,$ $\sigma _j^{-}=\left|
0_j\right\rangle \left\langle 1_j\right| ,$ and $\sigma _j^{+}=\left|
1_j\right\rangle \left\langle 0_j\right| ,$ with $\left| 0_j\right\rangle $
and $\left| 1_j\right\rangle $ ($j=1,2,...,n+1$) being the ground state and
excited level of the $j$th qubit. In the interaction picture with respect to
$H_0$, the above Hamiltonians $H_1$ and $H_2$ are rewritten respectively as
(assuming $\omega =\omega _0$)
\begin{eqnarray}
H_1 &=&\frac{\hbar \Omega }2\left( e^{i\varphi }S_{-}+e^{-i\varphi
}S_{+}\right) , \\
H_2 &=&\hbar g\left( e^{i\delta t}aS_{+}+e^{-i\delta t}a^{\dagger
}S_{-}\right) ,
\end{eqnarray}
where
\begin{equation}
\delta =\omega _0-\omega _c
\end{equation}
is the detuning between the $\left| 0\right\rangle \leftrightarrow \left|
1\right\rangle $ transition frequency $\omega _0$ of each qubit and the
frequency $\omega _c$ of the cavity mode.

We now consider two special cases: $\varphi =\pi $ and negative detuning $%
\delta <0$, as well as $\varphi =0$ and positive detuning $\delta >0.$ The
detailed discussion of these two cases will be given in the following
subsections III.A and III.B. The results from the unitary evolution,
obtained for these two special cases, will be employed by the two
alternative approaches (presented in section IV) for the gate implementation.

\begin{center}
\textbf{A. Case for pulse phase $\varphi =\pi $ and detuning $\delta <0$}
\end{center}

In this subsection, we consider the negative detuning case $\delta <0$ [Fig.
2(a)]. When $\varphi =\pi ,$ the above Hamiltonian (8) reduces to
\begin{equation}
H_1=-\frac{\hbar \Omega }2S_x,
\end{equation}
where
\begin{equation}
S_x=S_{-}+S_{+}=\sum_{j=1}^{n+1}\sigma _{x,j},
\end{equation}
with $\sigma _{x,j}=\sigma _j^{-}+\sigma _j^{+}.$ Performing the unitary
transformation $\left| \psi \left( t\right) \right\rangle =e^{-iH_1t}\left|
\psi ^{\prime }\left( t\right) \right\rangle ,$ we obtain
\begin{equation}
i\hbar \frac{d\left| \psi ^{\prime }\left( t\right) \right\rangle }{dt}=%
\overline{H}_2\left| \psi ^{\prime }\left( t\right) \right\rangle ,
\end{equation}
with
\begin{eqnarray}
\overline{H}_2 &=&\exp \left[ iH_1t/\hbar \right] H_2\exp \left[
-iH_1t/\hbar \right]  \nonumber \\
&=&\frac{\hbar g}2\left\{ e^{i\delta t}a\left[ S_x+\frac 12\left(
S_z-S_{-}+S_{+}\right) e^{-i\Omega t}\right. \right.  \nonumber \\
&&\left. \left. -\frac 12\left( S_z+S_{-}-S_{+}\right) e^{i\Omega t}\right]
\right\} +\text{H.c.},
\end{eqnarray}
Assuming that $\Omega \gg \left| \delta \right| ,g,$ we can neglect the
fast-oscillating terms [40-42]. Then the Hamiltonian (14) reduces to [40-42]
\begin{equation}
\overline{H}_2=\frac{\hbar g}2\left( e^{i\delta t}a+e^{-i\delta t}a^{\dagger
}\right) S_x.
\end{equation}

\begin{figure}[tbp]
\includegraphics[bb=114 309 487 528, width=8.0 cm, clip]{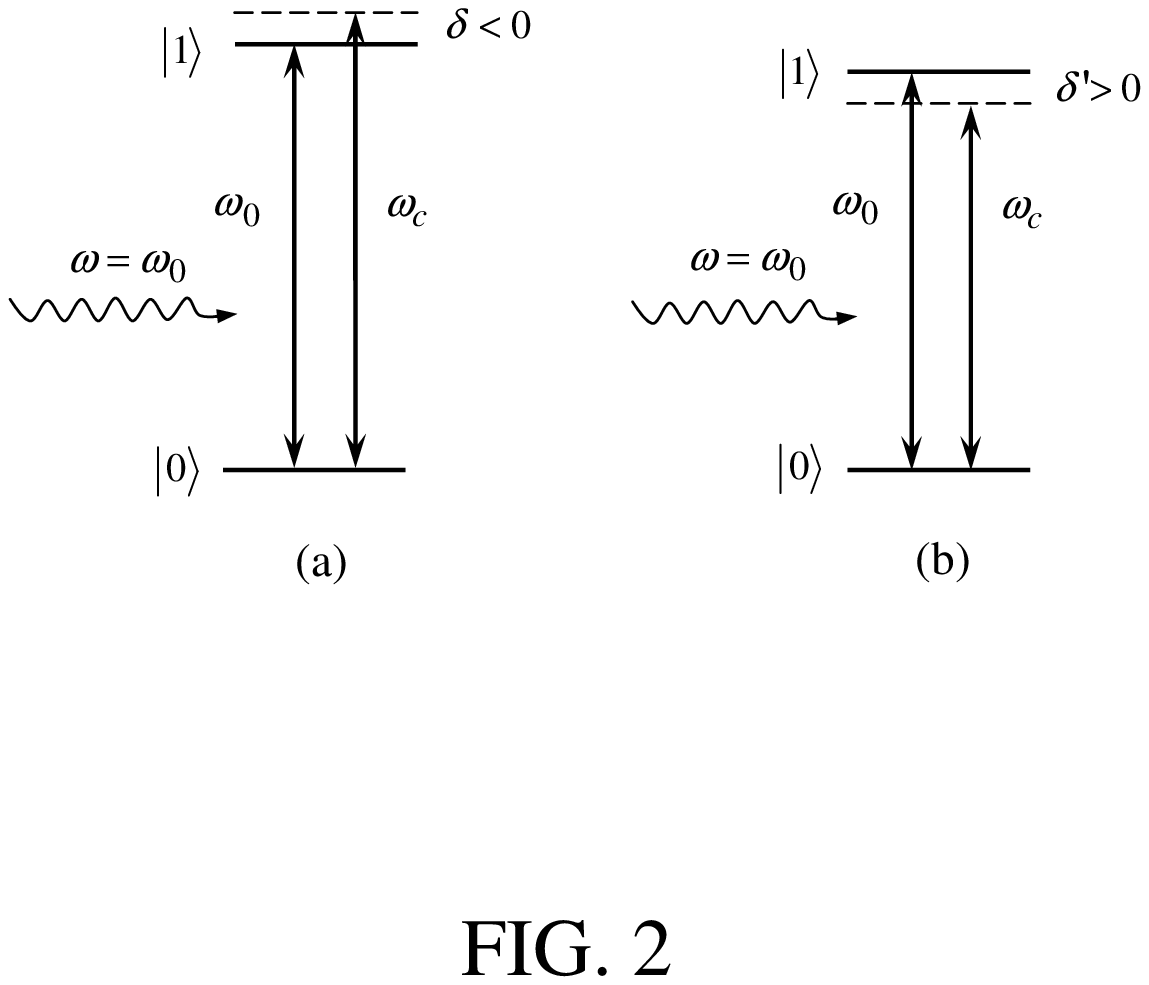} %
\vspace*{-0.18in}
\caption{Illustration of different detunings between the cavity mode
frequency $\omega _c$ and the qubit transition frequency $\omega _0$. Here, $%
\omega _0$, $\omega _c$, and $\omega $ are the qubit transition frequency,
the cavity mode frequency, and the pulse frequency, respectively. In (a),
the detuning $\delta =\omega _0-\omega _c<0.$ In (b), the detuning $%
\delta^{\prime } =\omega _0-\omega _c>0.$ To implement an NTCP gate, we
propose two alternative approaches (see section IV), each one with three
steps. The first step for either method requires a negative detuning ($%
\delta<0$), as shown in (a). For either method, the second step requires a
positive detuning ($\delta^{\prime}>0$), as shown in (b).}
\label{fig:2}
\end{figure}

The evolution operator for the Hamiltonian (15) can be written in the form
[19,40,43,44]
\begin{equation}
u\left( t\right) =e^{-iA\left( t\right) S_x^2}e^{-iB\left( t\right)
S_xa}e^{-iB^{*}\left( t\right) S_xa^{\dagger }},
\end{equation}
where
\begin{eqnarray}
B\left( t\right) &=&\frac g2\int_0^te^{i\delta t^{\prime }}dt^{\prime
}=\frac g{2i\delta }\left( e^{i\delta t}-1\right) ,  \nonumber \\
A\left( t\right) &=&i\int_0^tB\left( t^{\prime }\right) \frac{dB^{*}\left(
t^{\prime }\right) }{dt^{\prime }}dt^{\prime }  \nonumber \\
&=&\frac{g^2}{4\delta }\left[ t+\frac 1{i\delta }\left( e^{-i\delta
t}-1\right) \right] .
\end{eqnarray}
When the evolution time $t$ satisfies $t=\tau =2\pi /\left| \delta \right| ,$
we have $B\left( \tau \right) =0$ and $A\left( \tau \right) =g^2\tau /\left(
4\delta \right) $. Then we obtain
\begin{equation}
u\left( \tau \right) =\exp (i\lambda \tau S_x^2),
\end{equation}
where $\lambda =-g^2/\left( 4\delta \right) >0.$ The evolution operator of
the system (in the interaction picture with respect to $H_0$) is thus given
by
\begin{equation}
U\left( \tau \right) =e^{-iH_1\tau /\hbar }u\left( \tau \right) =e^{i\Omega
\tau S_x/2}e^{i\lambda \tau S_x^2}.
\end{equation}
In section IV, we propose two alternative approaches for the NTCP gate
implementation, each one involving three steps. The evolution operator $U$
in Eq. (19) will be needed for the first step for either one of the two
alternative approaches.

\begin{center}
\textbf{B. Case for pulse phase $\varphi =0$ and detuning $\delta >0$}
\end{center}

In the following, we consider the positive detuning case $\delta >0$. The
detuning $\delta $ can be adjusted from $\delta <0$ to $\delta >0$ by
changing either the qubit transition frequency $\omega _0$ or the cavity
mode frequency $\omega _c$, or both. In general, the qubit-cavity coupling
constant varies when the detuning changes. To distinguish this case ($%
\varphi =0$\textbf{\ }and\textbf{\ }$\delta >0$) from the previous case ($%
\varphi =\pi $\textbf{\ }and\textbf{\ }$\delta <0$), we replace the previous
notation $\Omega ,$ $\delta ,$ and $g$ by $\Omega ^{\prime },$ $\delta
^{\prime },$ and $g^{\prime }$, respectively [see Fig. 2(b)]. For
simplicity, we use the same symbols $\omega ,$ $\omega _0,$ and $\omega _c$
for the pulse frequency, the qubit transition frequency, and the cavity mode
frequency [Fig. 2(b)].

Suppose now that qubit $1$ is largely detuned (decoupled) from both the
cavity mode and the pulse. This can be achieved by adjusting the level
spacing of qubit $1$ (this tunability of energy levels can be achieved in
different types of qubits, e.g., solid-state qubits [30-34]), or by moving
qubit $1$ out of the cavity mode (e.g., trapped atoms [20,45,46]). Thus,
after dropping the terms related to qubit $1$ (i.e., the terms corresponding
to the index $j=1$) from the collective operators $S_{+}=\sum_{j=1}^{n+1}%
\sigma _j^{+}$ and $S_{-}=\sum_{j=1}^{n+1}\sigma _j^{-}$ in Hamiltonians (8)
and (9), and replacing $\Omega ,$ $\delta ,$ and $g$ (for the case of $%
\varphi =\pi $\textbf{\ }and\textbf{\ }$\delta <0$) with $\Omega ^{\prime },$
$\delta ^{\prime },$ and $g^{\prime }$ (for the case of $\varphi =0$\textbf{%
\ }and\textbf{\ }$\delta >0$), we obtain from the Hamiltonians (8) and (9)
(assuming $\omega =\omega _0$)
\begin{eqnarray}
H_1^{\prime } &=&\frac{\hbar \Omega ^{\prime }}2\left( e^{i\varphi
}S_{-}^{\prime }+e^{-i\varphi }S_{+}^{\prime }\right) , \\
H_2^{\prime } &=&\hbar g^{\prime }\left( e^{i\delta ^{\prime
}t}aS_{+}^{\prime }+e^{-i\delta ^{\prime }t}a^{\dagger }S_{-}^{\prime
}\right) ,
\end{eqnarray}
which are written in the interaction picture with respect to $H_0$ in
Eq.~(4). Here, $S_{-}^{\prime }$ and $S_{+}^{\prime }$ are the collective
operators for the $n$ qubits ($2,3,...,n+1$), which are given by

\begin{equation}
S_{-}^{\prime }=\sum_{j=2}^{n+1}\sigma _j^{-},\;S_{+}^{\prime
}=\sum_{j=2}^{n+1}\sigma _j^{+}.
\end{equation}

In addition, the detuning $\delta ^{\prime }$ is
\begin{equation}
\delta ^{\prime }=\omega _0-\omega _c>0.
\end{equation}

When choosing $\varphi =0,$ the Hamiltonian (20) reduces to
\begin{equation}
H_1^{\prime }=\frac{\hbar \Omega ^{\prime }}2S_x^{\prime },
\end{equation}
where
\begin{equation}
S_x^{\prime }=S_{-}^{\prime }+S_{+}^{\prime }=\sum_{j=2}^{n+1}\sigma _{x,j}.
\end{equation}
Note that the Hamiltonian~(24) has a form similar to Eq.~(11) and that the
Hamiltonian (21) has a form similar to Eq.~ (9).~Therefore, it is
straightforward to show that under the condition $\Omega ^{\prime }\gg
\delta ^{\prime },$ $g^{\prime },$ when $t=\tau ^{\prime }=2\pi /\delta
^{\prime },$ the evolution operator of the qubits (in the interaction
picture with respect to $H_0$) is
\begin{equation}
U^{\prime }\left( \tau ^{\prime }\right) =\exp (-i\Omega ^{\prime }\tau
^{\prime }S_x^{\prime }/2)\exp (-i\lambda ^{\prime }\tau ^{\prime
}S_x^{\prime 2}),
\end{equation}
where $\lambda ^{\prime }=g^{\prime 2}/\left( 4\delta ^{\prime }\right) >0.$
The evolution operator $U^{\prime }$ in Eq.~(26) will be needed for the
second step of either one of the two alternative approaches (presented in
section IV below) for implementing the NTCP gate.

One can see that the operator described by Eq.~(19) or Eq.~(26) does not
include the photon operator $a$ or $a^{\dagger }$ of the cavity mode.
Therefore, the cavity mode can be initially in an \textit{arbitrary} state
(e.g., in a vacuum state, a Fock state, a coherent state, or even a thermal
state).

For the gate realization, we will also need to have the qubits decoupled
from the cavity mode and apply a resonant pulse to each qubit. Suppose that
the Rabi frequency for the pulse applied to qubit $1$ is $\Omega _1$ while
the Rabi frequency for the pulses applied to qubits ($2,3,...,n+1$) is $%
\Omega _r.$ The initial phase for each pulse is $\varphi =0$. The free
Hamiltonian of the qubits and the cavity mode is the $H_0$ given in Eq. (4).
In the interaction picture with respect to $H_0$, the interaction
Hamiltonian for the qubit system and the pulses is given by
\begin{equation}
\widetilde{H}=\frac{\hbar \Omega _1}2\sigma _{x,1}+\frac{\hbar \Omega _r}%
2S_x^{\prime }.
\end{equation}
For an evolution time $\tau $, the time evolution operator for the
Hamiltonian (27) is
\begin{equation}
\widetilde{U}\left( \tau \right) =\exp \left[ -i\Omega _1\tau \sigma
_{x,1}/2\right] \exp \left[ -i\Omega _r\tau S_x^{\prime }/2\right] .
\end{equation}

The evolution operator $\widetilde{U}$ in Eq. (28) will be needed for the
third step for either one of the two alternative approaches below.

\begin{center}
\textbf{IV. IMPLEMENTATION OF THE NTCP GATE}
\end{center}

The goal of this section is to demonstrate how the NTCP gate can be realized
based on the unitary operators (19), (26), and (28) obtained above. We will
present two alternative methods for the gate implementation: one method
based on the adjustment of the qubit transition frequency and another method
which works mainly via the adjustment of the cavity mode frequency.

\begin{center}
\textbf{A. NTCP gate via adjusting the qubit transition frequency}
\end{center}

Let us consider $\left( n+1\right) $ qubits placed in a single-mode cavity
or coupled to a resonator. For this first method, the cavity mode frequency $%
\omega _c$ is kept fixed. The operations for the gate implementation, and
the unitary evolutions after each step of operation, are listed below:

\begin{figure}[tbp]
\includegraphics[bb=131 180 503 726, width=8.0 cm, clip]{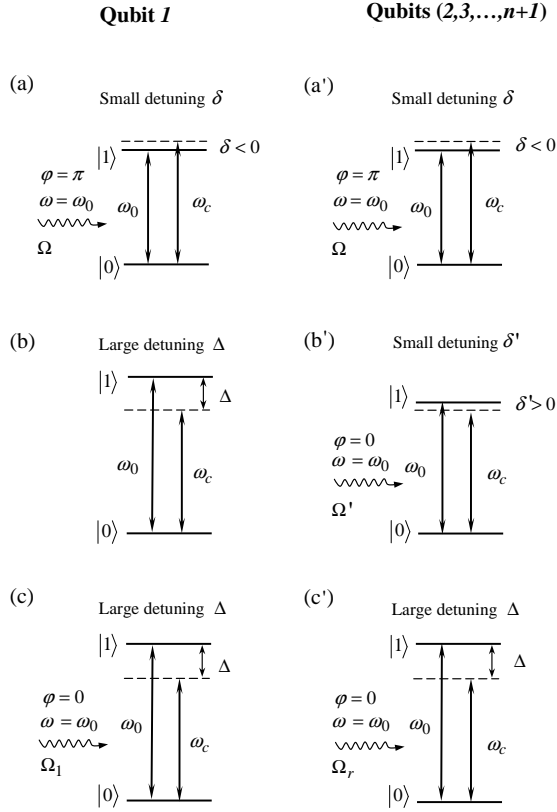} %
\vspace*{-0.18in}
\caption{Change of the qubit transition frequency $\omega _0$ (or the qubit
level spacings) of each qubit during a three-step NTCP gate. The three
figures on the left side corresponds to the control qubit $1$ while the
three figures on the right side corresponds to each of the target qubits ($%
2,3,...,n+1$). Figures (a) and (a$^{\prime }$), figures (b) and (b$^{\prime
} $), and figures\ (c) and (c$^{\prime }$) correspond to the operations of
step (i), step (ii), and step (iii), respectively. In each figure, the two
horizontal solid lines represent the qubit energy levels for the states $%
\left| 0\right\rangle $ and $\left| 1\right\rangle ;$ $\omega _0$ is the
qubit $\left| 0\right\rangle \leftrightarrow \left| 1\right\rangle $
transition frequency; $\omega _c$ is the cavity mode frequency; $\omega $ is
the pulse frequency; $\varphi $ is the initial phase of the pulse; and $%
\Omega $, $\Omega^{\prime },$ $\Omega _1$, and $\Omega _r$ are the Rabi
frequencies of various applied pulses. In addition, $\delta$ and $\delta
^{\prime }$ are the small detunings of the cavity mode with the $\left|
0\right\rangle \leftrightarrow \left| 1\right\rangle $ transition, which are
given by $\delta =\omega _0-\omega _c<0$ and $\delta ^{\prime }=\omega
_0-\omega _c>0$; while $\Delta=\omega _0-\omega _c$ represents the large
detuning of the cavity mode with the $\left| 0\right\rangle \leftrightarrow
\left| 1\right\rangle $ transition. Note that the cavity mode frequency $%
\omega _c$ is kept fixed during the entire process, but the qubit transition
frequency $\omega _0$ is adjusted to achieve a different detuning $\delta,
\delta^{\prime}$, or $\Delta$ for each step.}
\label{fig:3}
\end{figure}

Step (i): Apply a resonant pulse (with $\varphi =\pi $) to each qubit. The
pulse Rabi frequency is $\Omega .$ The cavity mode is coupled to each qubit
with a detuning $\delta <0$ [Fig. 3(a) and Fig. 3(a$^{\prime }$)]. This is
the case discussed in subsection III.A. Thus, the $U$ in Eq.~(19) is the
evolution operator for the qubit system for an interaction time $\tau =-2\pi
/\delta .$

Step\ (ii): Adjust the qubit transition frequency for qubits ($2,3,...,n+1$%
), such that the cavity mode is coupled to qubits ($2,3,...,n+1$) with a
detuning $\delta ^{\prime }>0$ [Fig. 3(b$^{\prime }$)]. Apply a resonant
pulse (with $\varphi =0$) to each of the target qubits ($2,3,...,n+1$) [Fig.
3(b$^{\prime }$)]. The pulse Rabi frequency is now $\Omega ^{\prime }.$ In
addition, adjust the transition frequency of qubit $1$ such that qubit $1$
is decoupled from the cavity mode and the pulses applied to qubits ($%
2,3,...,n+1$) [Fig. 3(b)]. It can be seen that this is the case discussed in
subsection III.B. Thus, the $U^{\prime }$ in Eq.~(26) is the evolution
operator for the qubit system for an interaction time $\tau ^{\prime }=2\pi
/\delta ^{\prime }.$

The combined time evolution operator, after the above two steps, is
\begin{eqnarray}
U\left( \tau +\tau ^{\prime }\right) &=&U^{\prime }\left( \tau ^{\prime
}\right) U\left( \tau \right)  \nonumber \\
\ &=&e^{-i\Omega ^{\prime }\tau ^{\prime }S_x^{\prime }/2}e^{-i\lambda
^{\prime }\tau ^{\prime }S_x^{\prime 2}}e^{i\Omega \tau S_x/2}e^{i\lambda
\tau S_x^2}  \nonumber \\
\ &=&\exp \left[ i\Omega \tau \left( S_x-S_x^{\prime }\right) /2\right] \exp
\left[ i\lambda \tau \left( S_x^2-S_x^{\prime 2}\right) \right] .  \nonumber
\\
&&
\end{eqnarray}
In the last line of Eq.~(29), we assumed $\Omega \tau =\Omega ^{\prime }\tau
^{\prime }$ (i.e., $-\Omega /\delta =\Omega ^{\prime }/\delta ^{\prime }$)
and $\lambda \tau =\lambda ^{\prime }\tau ^{\prime }$ (i.e., $-g/\delta
=g^{\prime }/\delta ^{\prime }$), which can be achieved by adjusting the
detunings $\delta $ and $\delta ^{\prime }$ (i.e., changing the qubit
transition frequency) as well as the Rabi frequencies $\Omega $ and $\Omega
^{\prime }$ (i.e., changing the intensity/amplitude of the pulses). Note
that $S_x-S_x^{\prime }=\sigma _{x,1}$ and $S_x^2-S_x^{\prime 2}=I+2\sigma
_{x,1}S_x^{\prime },$ where $I$ is the identity operator for qubit $1.$
Thus, Eq.~(29) can be written as
\begin{equation}
U\left( \tau +\tau ^{\prime }\right) =\exp \left[ i\Omega \tau \sigma
_{x,1}/2\right] \exp \left[ i2\lambda \tau \sigma _{x,1}S_x^{\prime }\right]
.
\end{equation}

Step: (iii) Leave the transition frequency unchanged for qubit $1$ [Fig.
~3(c)] but adjust the transition frequency of qubits ($2,3,...,n+1$)
[Fig.~3(c$^{\prime }$)], such that the cavity mode is largely detuned
(decoupled) from each qubit. In addition, apply a resonant pulse (with $%
\varphi =0$) to each qubit. The Rabi frequency for the pulse applied to
qubit $1$ is now $\Omega _1$ [Fig. 3(c)] while the Rabi frequency of the
pulse applied to qubits ($2,3,...,n+1$) is $\Omega _r$ [Fig. 3(c$^{\prime }$%
)]. In this case, the interaction Hamiltonian for the qubit system and the
pulses is given by Eq.~(27) and thus the time evolution operator is the $%
\widetilde{U}$ in Eq.~(28) for an evolution time $\tau $ given above.

It can be seen that after the above three-step operation, the joint
time-evolution operator of the qubit system is
\begin{eqnarray}
U\left( 2\tau +\tau ^{\prime }\right) &=&\widetilde{U}\left( \tau \right)
U\left( \tau +\tau ^{\prime }\right)  \nonumber \\
\ &=&e^{-i(\Omega _1-\Omega )\tau \sigma _{x,1}/2}e^{-i\Omega _r\tau
S_x^{\prime }/2}e^{i2\lambda \tau \sigma _{x,1}S_x^{\prime }}.  \nonumber \\
&&
\end{eqnarray}
Under the following condition
\begin{eqnarray}
\Omega _1 &=&4\lambda n+\Omega =-ng^2/\delta +\Omega ,\text{ }  \nonumber \\
\Omega _r &=&4\lambda =-g^2/\delta,
\end{eqnarray}
(which can be obtained by adjusting the Rabi frequencies $\Omega ,$ $\Omega
_1$ and $\Omega _r$), Eq. (31) can be then written as
\begin{equation}
U\left( 2\tau +\tau ^{\prime }\right) =\prod_{j=2}^{n+1}U_p\left( 1,j\right),
\end{equation}
with
\begin{equation}
U_p\left( 1,j\right) =\exp \left[ -i2\lambda \tau \left( \sigma
_{x,1}+\sigma _{x,j}-\sigma _{x,1}\sigma _{x,j}\right) \right] .
\end{equation}
It can be easily shown that for the qubit pair ($1,j$), we have
\begin{eqnarray}
U_p\left( 1,j\right) \left| +_1\right\rangle \left| +_j\right\rangle
&=&\left| +_1\right\rangle \left| +_j\right\rangle ,  \nonumber \\
U_p\left( 1,j\right) \left| +_1\right\rangle \left| -_j\right\rangle
&=&\left| +_1\right\rangle \left| -_j\right\rangle ,  \nonumber \\
U_p\left( 1,j\right) \left| -_1\right\rangle \left| +_j\right\rangle
&=&\left| -_1\right\rangle \left| +_j\right\rangle ,  \nonumber \\
U_p\left( 1,j\right) \left| -_1\right\rangle \left| -_j\right\rangle &=&\exp
(i8\lambda \tau )\left| -_1\right\rangle \left| -_j\right\rangle ,
\end{eqnarray}
where an overall phase factor $\exp (-i2\lambda \tau )$ is omitted. Here and
below, $\left| +_1\right\rangle =\left( \left| 0_1\right\rangle +\left|
1_1\right\rangle \right) /\sqrt{2},$ and $\left| -_1\right\rangle =\left(
\left| 0_1\right\rangle -\left| 1_1\right\rangle \right) /\sqrt{2},$ are the
two eigenstates of the Pauli operator $\sigma _{x,1}$ for qubit $1;$ while $%
\left| +_j\right\rangle =\left( \left| 0_j\right\rangle +\left|
1_j\right\rangle \right) /\sqrt{2},$ and $\left| -_j\right\rangle =\left(
\left| 0_j\right\rangle -\left| 1_j\right\rangle \right) /\sqrt{2},$ are the
two eigenstates of the Pauli operator $\sigma _{x,j}$ for qubit $j$ ($%
j=2,3,...,$ or $n+1$). By setting $8\lambda \tau =\left( 2k+1\right) \pi ,$
i.e.,
\begin{equation}
4g^2/\delta ^2=2k+1
\end{equation}
($k$ is an integer), we obtain from Eq. (35)
\begin{eqnarray}
U_p\left( 1,j\right) \left| +_1\right\rangle \left| +_j\right\rangle
&=&\left| +_1\right\rangle \left| +_j\right\rangle ,  \nonumber \\
U_p\left( 1,j\right) \left| +_1\right\rangle \left| -_j\right\rangle
&=&\left| +_1\right\rangle \left| -_j\right\rangle ,  \nonumber \\
U_p\left( 1,j\right) \left| -_1\right\rangle \left| +_j\right\rangle
&=&\left| -_1\right\rangle \left| +_j\right\rangle ,  \nonumber \\
U_p\left( 1,j\right) \left| -_1\right\rangle \left| -_j\right\rangle
&=&-\left| -_1\right\rangle \left| -_j\right\rangle ,
\end{eqnarray}
which shows that a quantum phase gate described by
\begin{equation}
U_p\left( 1,j\right) =I_j-2\left| -_1-_j\right\rangle \left\langle
-_1-_j\right| ,
\end{equation}
is achieved for the qubit pair ($1,j$). Here, $I_j$ is the identity operator
for the qubit pair ($1,j$), which is given by $I_j=\sum_{kl}\left|
k_1l_j\right\rangle \left\langle k_1l_j\right| $ with $k,l\in \left\{
+,-\right\} .$ Note that the condition (36) can be achieved by adjusting the
detuning $\delta $ (i.e., via changing the qubit transition frequency $%
\omega _0$).

Combining Eqs.~(33) and (38), we finally obtain
\begin{equation}
U\left( 2\tau +\tau ^{\prime }\right) =\prod_{j=2}^{n+1}\left( I_j-2\left|
-_1-_j\right\rangle \left\langle -_1-_j\right| \right) ,
\end{equation}
which demonstrates that $n$ two-qubit CP gates are simultaneously performed
on the qubit pairs ($1,2$), ($1,3$),..., and ($1,n+1$), respectively. Note
that each qubit pair contains the same control qubit (qubit $1$) and a
different target qubit (either qubit $2,$ $3,$ $...,$ or $n+1$). Hence, an
NTCP gate with $n$ target qubits ($2,3,...,n+1$) and one control qubit
(qubit 1) is obtained after the above three-step process.

From the description above, one can see that the method presented here has
an advantage: it does not require the adjustment of the cavity-mode
frequency.

\begin{center}
\textbf{B. NTCP gate via mainly adjusting the cavity-mode frequency}
\end{center}

\begin{figure}[tbp]
\includegraphics[bb=131 176 500 723, width=8.0 cm, clip]{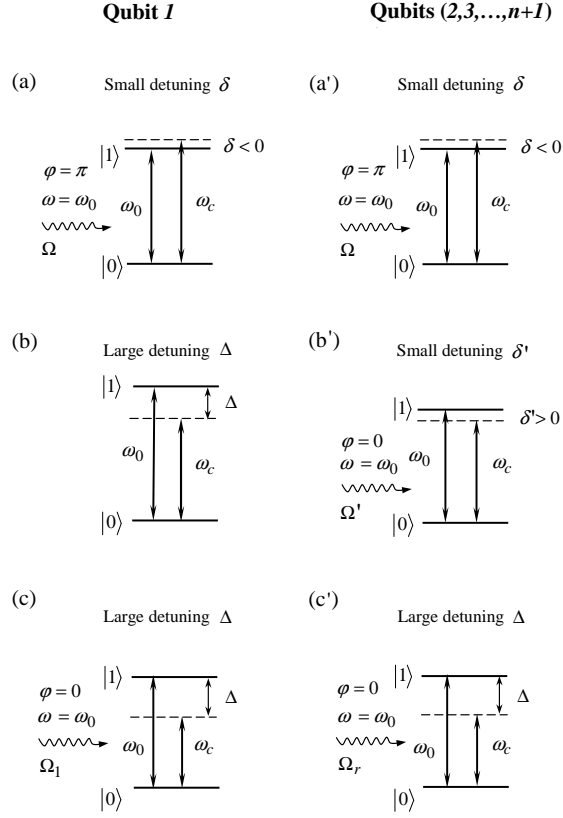} %
\vspace*{-0.18in}
\caption{Change of the cavity mode frequency $\omega _c$ and the transition
frequency $\omega _0$ (or the level spacings) of qubit $1$ during a
three-step NTCP gate. This second method is an alternative way to implement
the NTCP gate, which is different from the steps shown in Fig. 3. Note that
the transition frequency $\omega _0$ for qubits ($2,3,...,n+1$) remains
unchanged during this method. The three figures on the left side correspond
to the control qubit $1,$ which (from top to bottom) are for the operations
of step (i), step (ii), and step (iii), respectively. The three figures on
the right side correspond to each of the qubits ($2,3,...,n+1$), which (from
top to bottom) are respectively for the operations of step (i), step (ii),
and step (iii). In each figure, the two horizontal solid lines represent the
qubit levels $\left| 0\right\rangle $ and $\left| 1\right\rangle ;$ $\omega
_0$ is the qubit transition frequency; $\omega _c$ is the cavity mode
frequency; $\omega $ is the pulse frequency; $\varphi $ is the initial phase
of the pulse; and $\Omega $, $\Omega ^{\prime },$ $\Omega _1$, and $\Omega
_r $ are the Rabi frequencies of the pulses (for various steps). In
addition, $\delta$ and $\delta ^{\prime } $ are the small detunings of the
cavity mode with the $\left| 0\right\rangle \leftrightarrow \left|
1\right\rangle $ transition, which are given by $\delta =\omega _0-\omega
_c<0$ and $\delta ^{\prime }=\omega _0-\omega _c>0$; while $\Delta=\omega
_0-\omega _c$ represents the large detuning of the cavity mode with the $%
\left| 0\right\rangle \leftrightarrow \left| 1\right\rangle $ transition.}
\label{fig:4}
\end{figure}

In this subsection, we present a different way for realizing the NTCP gate,
which is mainly based on the adjustment of the cavity mode frequency $\omega
_c$.

Let us consider $(n+1)$ qubits placed in a single-mode cavity or coupled to
a resonator. Note that now the transition frequency $\omega _0$ of the $n$
target qubits ($2,3,...,n+1$) is kept fixed during the following entire
process. The operations for the gate implementation and the unitary
evolutions after each step are listed below:

Step (i): Apply a resonant pulse (with $\varphi =\pi $) to each qubit. The
pulse Rabi frequency is $\Omega .$ The cavity mode is coupled to each qubit
with a detuning $\delta <0$ [Fig. 4(a) and Fig. 4(a$^{\prime }$)]. It can be
seen that this is the case discussed in subsection III.A. Thus, the $U$ in
Eq.~(19) is the evolution operator for the qubit system for an interaction
time $\tau =-2\pi /\delta .$

Step (ii): Leave the transition frequency $\omega _0$ of qubits ($%
2,3,...,n+1 $) unchanged while adjusting the cavity mode frequency $\omega
_c $. The cavity mode is coupled to qubits ($2,3,...,n+1$) with a detuning $%
\delta ^{\prime }>0$ [Fig. 4(b$^{\prime }$)]. Apply a resonant pulse (with $%
\varphi =0$) to each of qubits ($2,3,...,n+1$) [Fig. 4(b$^{\prime }$)]. The
pulse Rabi frequency is now $\Omega ^{\prime }.$ In addition, adjust the
transition frequency of qubit $1,$ such that qubit $1$ is largely detuned
(decoupled) from the cavity mode as well as the pulses applied to qubits ($%
2,3,...,n+1$) [Fig. 4(b)]. One can see that this is the case discussed in
subsection III.B. Hence, the $U^{\prime }$ in Eq.~ (26) is the evolution
operator for the qubit system for an interaction time $\tau ^{\prime }=2\pi
/\delta ^{\prime }.$

Step (iii): Leave the transition frequency $\omega _0$ of qubits ($%
2,3,...,n+1$) unchanged while adjusting the transition frequency of the
control qubit $1$ back to its original setting in step (i) [Fig. 4(c)].
Adjust the cavity mode frequency $\omega _c$, such that the cavity mode is
largely detuned (decoupled) from each qubit [Fig. 4(c) and Fig. 4(c$^{\prime
}$)]. In addition, apply a resonant pulse (with $\varphi =0$) to each qubit.
The Rabi frequency for the pulse applied to qubit $1$ is now $\Omega _1$
[Fig. 4(c)] while the Rabi frequency of the pulses applied to qubits ($%
2,3,...,n+1$) is now $\Omega _r$ [Fig. 4(c$^{\prime }$)]. Therefore, the
interaction Hamiltonian for the qubits and the pulses is $H_1$ in Eq.~(27)
and thus the time evolution operator is the $\widetilde{U}$ in Eq.~(28) for
an evolution time $\tau $ given above.

Note that the time evolution operators $U,$ $U^{\prime },$ and $\widetilde{U}
$, obtained from each step discussed in this subsection, are the same as
those obtained from each step of operations in the previous subsection.
Hence, it is clear that the NTCP gate can be implemented after this
three-step process.

From the description above, it can be seen that the transition frequency $%
\omega _0$ for each one of the $n$ target qubits ($2,3,...,n+1$) remains
unchanged during the entire operation. Thus, adjusting the level spacings
for the $n$ target qubits ($2,3,...,n+1$) is not required by the method
presented in this subsection. What one needs to do is to adjust the cavity
mode frequency and the level spacings of the control qubit $1$.

\begin{center}
\textbf{C. Discussion}
\end{center}

We have presented two alternative methods for implementing an NTCP gate.
Note that when coupled to a cavity (or resonator) mode and driven by a
classical pulse, physical qubit systems (such as atoms, quantum dots, and
superconducting qubits) have the same type of interaction described by the
Hamiltonian (3) or a Hamiltonian having a similar form to Eq.~(3), from
which the four Hamiltonians (8), (9), (20), and (21), i.e., the key elements
for the proposed NTCP gate implementation, are available. Therefore, the two
alternative methods above are quite general, which can be applied to
implementing the NTCP gate with superconducting qubits, quantum dots, or
trapped atomic qubits.

As shown above, the first method is based on the adjustment of the qubit
level spacings while the second method works mainly via the adjustment of
the cavity mode frequency. For solid-state qubits, the qubit level spacings
can be rapidly adjusted (e.g., in 1 $\sim$ 2 nanosecond timescale for
superconducting qubits [47]). And, for superconducting resonators, the
resonator frequency can be fast tuned (e.g., in less than a few nanoseconds
for a superconducting transmission line resonator [36]).

In the appendix, we provide guidelines on protecting qubits from leaking out
of the computational subspace in the presence of more qubit levels. As
discussed there, as long as the large detuning of the cavity mode with the
transition between the irrelevant energy levels can be met, the leakage out
of the computational subspace induced due to the cavity mode can be
suppressed. In addition, we should mention that since the applied pulses are
\textit{resonant} with the $\left| 0\right\rangle \leftrightarrow \left|
1\right\rangle $ transition, the coupling of the qubit level $\left|
0\right\rangle $ or $\left| 1\right\rangle $ with other levels, induced by
the pulses, is negligibly small. Therefore, the leak-out of the computation
subspace, due to the application of the pulses, can be neglected.

As discussed above, during the gate operation for either of the two methods,
decoupling of the qubits from the cavity was made by adjusting the qubit
frequency $\omega _0$ or the cavity frequency $\omega _c$. This particularly
applies to solid state qubits because the locations of solid-state qubits
(such as superconducting qubits and quantum dots) are fixed once they are
built in a cavity or resonator. However, it is noted that for trapped atomic
qubits, decoupling of qubits from the cavity can be made by just moving
atoms out of the cavity and thus adjustment of the level spacings of atoms
or the cavity frequency is not needed to have atoms decoupled (largely
detuned) from the cavity mode (e.g., as shown in Section VI).

For both step (i) and step (ii) of the two methods above, it was assumed
that the detuning $\delta$ ($\delta^{\prime }$) is much smaller than the
qubit Rabi frequency $\Omega$ ($\Omega^{\prime }$) and that the Rabi
frequencies are equal for all the qubits. In reality, the Rabi frequency for
each qubit might not be identical, and the effect of the small Rabi
frequency deviations on the gate performance may be magnified when the
operation time is much longer than the inverse Rabi frequency. Hence, to
improve the gate performance, the Rabi frequency deviations should be as
small as possible, which could be achieved by adjusting the intensity of the
pulses applied to the qubits.

We should mention that when the mean photon number $\overline{n}$ of the
cavity field satisfies $\overline{n}\leq 1$, the multi-photon process can be
neglected. In the case that the condition $\overline{n}\leq 1$ does not
meet, one can adjust the qubit level spacings or choose the qubit level
structures appropriately such that the multi-photon process is negligible in
the process of quantum operations.

Before closing this section, it should be mentioned that the present method
is based on an effective Hamiltonian
\begin{equation}
H_{\mathrm{eff}}={\small \sum_{j=2}^{n+1}}H_{1j}=2\hbar \lambda
\sum_{j=2}^{n+1}\left( \sigma _{x,1}+\sigma _{x,j}-\sigma _{x,1}\sigma
_{x,j}\right) ,
\end{equation}
which can be found from Eqs.~(33) and (34). One can see that this
Hamiltonian contains the interaction terms between the control qubit (qubit $%
1$) and each target qubit, but does not include the interaction terms
between any two target qubits.~Note that each term $H_{1j}$ in Eq.~(40) acts
on a different target qubit, with the same control qubit, and that any two
terms $H_{1j}$ for different $j$'s commute with each other. Therefore, the $%
n $ two-qubit controlled-phase gates forming the NTCP gate can be \textit{%
simultaneously} performed on the qubit pairs ($1,2$), ($1,3$),..., and ($%
1,n+1$), respectively.

\begin{center}
\textbf{V. REALIZING THE NTCP GATE WITH SUPERCONDUCTING QUBITS COUPLED TO A
RESONATOR}
\end{center}

The methods presented above for implementing the NTCP gate are based on the
four Hamiltonians (8), (9), (20) and (21). In this section, we show how
these Hamiltonians can be obtained for the superconducting charge qubits
coupled to a resonator. We will then show how to apply the first method
above to implement the NTCP gate with $\left( n+1\right) $ charge qubits
selected from $N$ charge qubits coupled to a resonator ($1<n<N$). A
discussion on the experimental feasibility will be given later.

\begin{center}
\textbf{A. Hamiltonians}
\end{center}

The superconducting charge qubit considered here, as shown in Fig.~5(a),
consists of a small superconducting box with excess Cooper-pair charges,
connected to a symmetric superconducting quantum interference device (SQUID)
with capacitance $C_{J0}$ and Josephson coupling energy $E_{J0}.$ In the
charge regime $\Delta ~\gg ~E_c\gg ~E_{J0}\gg ~k_BT$ (here,~$k_B,$~$\Delta ,$%
~$E_c,$~and~$T$ are the Boltzmann constant, superconducting energy gap,
charging energy, and temperature, respectively), only two charge states, $%
n=0 $ and $n=1$, are important for the dynamics of the system, and thus this
device~[e.g., 30-32] behaves as a two-level system. Here, we denote the two
charge states $n=0$ and $n=1$ as the two eigenstates $\left| +\right\rangle $
and $\left| -\right\rangle $ of the spin operator $\widetilde{\sigma }_z.$
The reason for not using the eigenstates $\left| 0\right\rangle $ and $%
\left| 1\right\rangle $ of $\sigma _z$ to represent the two charge states is
that: in order to obtain the four Hamiltonians (8), (9), (20), and (21), in
the following we will need to perform a basis transformation from the $%
\widetilde{\sigma }_z$ basis $\left\{ \left| +\right\rangle ,\left|
-\right\rangle \right\} $ to the $\sigma _z$ basis $\left\{ \left|
0\right\rangle ,\left| 1\right\rangle \right\} .$ The Hamiltonian describing
the qubit is given by [30,31]
\begin{equation}
H_q=-2E_c(1-2n_g)\widetilde{\sigma }_z-E_J\left( \Phi \right) \widetilde{%
\sigma }_x,
\end{equation}
where $n_g=C_gV_g/\left( 2e\right) $, $E_c=e^2/(2C_g+4C_{J0}),$ $E_J\left(
\Phi \right) =2E_{J0}\cos \left( \pi \Phi /\Phi _0\right) $, $\widetilde{%
\sigma }_z=\left| +\right\rangle \left\langle +\right| -\left|
-\right\rangle \left\langle -\right| $ and $\widetilde{\sigma }_x=\left|
+\right\rangle \left\langle -\right| +\left| -\right\rangle \left\langle
+\right| .$ Here, $C_g$ is the gate capacitance, $V_g$ is the gate voltage, $%
\Phi $ is the external magnetic flux applied to the SQUID loop, $\Phi
_0=h/2e $ is the flux quantum, and $E_J\left( \Phi \right) $ is the
effective Josephson coupling energy. We assume that $V_g=V_g^{\mathrm{dc}%
}+V_g^{\mathrm{ac}}+V_g^{\mathrm{qu}},$ where $V_g^{\mathrm{dc}}$ ($V_g^{%
\mathrm{ac}}$) is the dc (ac) part of the gate voltage and $V_g^{\mathrm{qu}%
} $ is the quantum part of the gate voltage, which is caused by the electric
field of the resonator mode when the qubit is coupled to a resonator.
Correspondingly, we have
\begin{equation}
n_g=n_g^{\mathrm{dc}}+n_g^{\mathrm{ac}}+n_g^{\mathrm{qu}},
\end{equation}
where $n_g^{\mathrm{dc}}=C_gV_g^{\mathrm{dc}}/\left( 2e\right) ,$ $n_g^{%
\mathrm{ac}}=C_gV_g^{\mathrm{ac}}/\left( 2e\right) ,$ and $n_g^{\mathrm{qu}%
}=C_gV_g^{\mathrm{qu}}/\left( 2e\right) .$ By inserting Eq. (42) into Eq.
(41), we obtain the following Hamiltonian for the qubit-cavity system
\begin{eqnarray}
H &=&E_z\widetilde{\sigma }_z-E_J\left( \Phi \right) \widetilde{\sigma }%
_x+\hbar \omega _ca^{\dagger }a  \nonumber \\
&&+4E_cn_g^{\mathrm{ac}}\widetilde{\sigma }_z+4E_cn_g^{\mathrm{qu}}%
\widetilde{\sigma }_z,
\end{eqnarray}
where $E_z=-2E_c(1-2n_g^{\mathrm{dc}}).$ When $V_g^{\mathrm{ac}}=V_0\cos
\left( \omega t+\varphi \right) $ and $V_g^{\mathrm{qu}}=V_0^{\mathrm{qu}%
}\left( a+a^{\dagger }\right) $, the Hamiltonian (43) becomes [11,48]
\begin{eqnarray}
H &=&E_z\widetilde{\sigma }_z-E_J\left( \Phi \right) \widetilde{\sigma }%
_x+\hbar \omega _ca^{\dagger }a  \nonumber \\
&&+\hbar \Omega \cos \left( \omega t+\varphi \right) \widetilde{\sigma }%
_z+\hbar g\left( a+a^{{}\dagger }\right) \widetilde{\sigma }_z,
\end{eqnarray}
where $\Omega =2E_cC_gV_0/\left( \hbar e\right) $ is the Rabi frequency of
the ac gate voltage and $g=2E_cC_gV_0^{\mathrm{qu}}/\left( \hbar e\right) $
is the coupling constant between the charge qubit and the resonator mode.

\begin{figure}[tbp]
\includegraphics[bb=66 313 550 694, width=8.6 cm, clip]{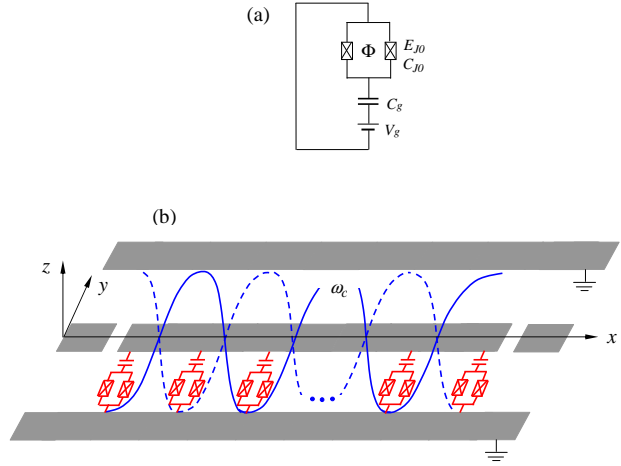} %
\vspace*{-0.18in}
\caption{(a) (Color online) Schematic diagram of a superconducting charge
qubit. Here, $E_{J0}$ is the Josephson coupling energy, $C_{J0}$ is the
Josephson capacitance, $C_g$ is the gate capacitance, $V_g$ is the gate
voltage, and $\Phi$ is the external magnetic flux applied to the SQUID loop
through a control line (not shown). (b) $N$ superconducting qubits (red
squares) are placed into a (grey) quasi-one dimensional transmission line
resonator. Each qubit is placed at an antinode of the electric field,
yielding a strong coupling between the qubit and the resonator mode. The two
blue curves represent the standing-wave electric field, along the $y$%
-direction. A subset of qubits from the $N$ qubits, selected for the gate,
are coupled to each other via the resonator mode; while the remaining
qubits, which are not controlled by the gate, are decoupled from the
resonator mode by setting their $\Phi=\Phi_0/2,~V_g^{\mathrm{dc}}=e/C_g,$%
~and $V_g^{\mathrm{ac}}=0$ to have their free Hamiltonian equal to zero.}
\label{fig:5}
\end{figure}

Let us now consider $N$ identical charge qubits coupled to a single-mode
resonator [Fig. 5(b)]. One can select a subset of qubits for the gate, while
the remaining qubits, which are not controlled by the gate, are decoupled
from the resonator mode by setting their $\Phi =\Phi _0/2,~V_g^{\mathrm{dc}%
}=e/C_g,$~and $V_g^{\mathrm{ac}}=0$ to have their free Hamiltonian equal to
zero. Without loss of generality, we assume that the set of qubits selected
for the gate are the $\left( n+1\right) $ qubits labelled by 1,~2,~...,~and~$%
n+1$ (here, $1<n<N$). From the discussion above, it can be seen that the
Hamiltonian for the $\left( n+1\right) $ qubits and the resonator mode is
\begin{eqnarray}
H &=&E_z\widetilde{S}_z-E_J\left( \Phi \right) \widetilde{S}_x+\hbar \omega
_ca^{\dagger }a  \nonumber \\
&&+\hbar \Omega \cos \left( \omega t+\varphi \right) \widetilde{S}_z+\hbar
g\left( a+a^{\dagger }\right) \widetilde{S}_z,
\end{eqnarray}
where $\widetilde{S}_z=\sum_{j=1}^{n+1}\widetilde{\sigma }_{z,j}$ and $%
\widetilde{S}_x=\sum_{j=1}^{n+1}\widetilde{\sigma }_{x,j},$ with $\widetilde{%
\sigma }_{z,j}=\left| +_j\right\rangle \left\langle +_j\right| -\left|
-_j\right\rangle \left\langle -_j\right| $ and $\widetilde{\sigma }%
_{x,j}=\left| +_j\right\rangle \left\langle -_j\right| +\left|
-_j\right\rangle \left\langle +_j\right| .$ By setting $E_z=0$ (i.e., $n_g^{%
\mathrm{dc}}=1/2$) for each qubit and defining $\omega _0/2=E_J\left( \Phi
\right) /\hbar $, the Hamiltonian (45) reduces to
\begin{eqnarray}
H &=&-\frac{\hbar \omega _0}2\widetilde{S}_x+\hbar \omega _ca^{\dagger }a
\nonumber \\
&&+\hbar \Omega \cos \left( \omega t+\varphi \right) \widetilde{S}_z+\hbar
g\left( a+a^{\dagger }\right) \widetilde{S}_z.
\end{eqnarray}
Define now the new qubit basis $\left| 0_j\right\rangle =\left( \left|
+_j\right\rangle +\left| -_j\right\rangle \right) /\sqrt{2}$ and $\left|
1_j\right\rangle =\left( \left| +_j\right\rangle -\left| -_j\right\rangle
\right) /\sqrt{2},$ i.e., perform a basis transformation from the $%
\widetilde{\sigma }_{z,j}$ basis $\left\{ \left| +_j\right\rangle ,\left|
-_j\right\rangle \right\} $ to the $\sigma _{z,j}$ basis $\left\{ \left|
0_j\right\rangle ,\left| 1_j\right\rangle \right\} $ for the $j$th qubit.
Thus, in the new basis, the Hamiltonian (46) becomes
\begin{equation}
H=H_0+H_1+H_2,
\end{equation}
where
\begin{eqnarray}
H_0 &=&-\frac{\hbar \omega _0}2S_z+\hbar \omega _ca^{\dagger }a, \\
H_1 &=&\hbar \Omega \cos \left( \omega t+\varphi \right) S_x, \\
H_2 &=&\hbar g\left( a+a^{\dagger }\right) S_x.
\end{eqnarray}
Here, the collective operators $S_z$ and $S_x$ are the same as those given
in Eq.~(7) and Eq.~(12). In the interaction picture with respect to $H_0,$
we obtain from Eqs.~(49) and (50) (under the rotating-wave approximation and
assuming $\omega =\omega _0$)

\begin{eqnarray}
H_1 &=&\frac{\hbar \Omega }2\left( e^{i\varphi }S_{-}+e^{-i\varphi
}S_{+}\right) , \\
H_2 &=&\hbar g\left( e^{i\delta t}aS_{+}+e^{-i\delta t}a^{\dagger
}S_{-}\right) ,
\end{eqnarray}
where the collective operators $S_{-}$ and $S_{+}$ are the same as those
given in Eq.~(7), and $\delta $ $=\omega _0-\omega _c<0.$

Note that $\omega _0=4E_{J0}\cos \left( \pi \Phi /\Phi _0\right) /\hbar $.
Hence, the qubit transition frequency $\omega _0$ can be adjusted by
changing the external magnetic flux $\Phi $ applied to the SQUID loop of the
charge qubit.

We now turn off the ac gate voltage applied to the charge qubit $1$ (i.e.,
setting $V_g^{\mathrm{ac}}=0$ for the charge qubit $1$) and adjust the
transition frequency $\omega _0$ of the charge qubit $1$ to have qubit $1$
decoupled (largely detuned) from the resonator mode. In this way, we can
drop the terms corresponding to the index $j=1$ from the collective
operators $S_{+}=\sum_{j=1}^{n+1}\sigma _j^{+}$ and $S_{-}=\sum_{j=1}^{n+1}%
\sigma _j^{-}$ involved in Hamiltonians (51) and (52). In addition, adjust
either the transition frequency $\omega _0$ of qubits ($2,3,...,n+1$) or the
resonator frequency $\omega _c$, to achieve a detuning $\delta ^{\prime
}=\omega _0-\omega _c>0,$ and set an ac gate voltage $V_g^{\mathrm{ac}%
}=V_0^{\prime }\cos \left( \omega t+\varphi \right) $ (with $\omega =\omega
_0$) for each of qubits ($2,3,...,n+1$). The Rabi frequency $\Omega ^{\prime
}$ for each ac gate voltage (i.e., the pulse) is given by $\Omega ^{\prime
}=2E_cC_gV_0^{\prime }/\left( \hbar e\right) .$ After replacing $\Omega ,$ $%
\delta ,$ and $g,$ with $\Omega ^{\prime },$ $\delta ^{\prime },$ and $%
g^{\prime }$, respectively; we can obtain from Eqs.~(51) and (52)
\begin{eqnarray}
H_1^{\prime } &=&\frac{\hbar \Omega ^{\prime }}2\left( e^{i\varphi
}S_{-}^{\prime }+e^{-i\varphi }S_{+}^{\prime }\right) , \\
H_2^{\prime } &=&\hbar g^{\prime }\left( e^{i\delta ^{\prime
}t}aS_{+}^{\prime }+e^{-i\delta ^{\prime }t}a^{\dagger }S_{-}^{\prime
}\right) ,
\end{eqnarray}
which are written in the interaction picture with respect to $H_0$ in
Eq.~(48). Here, the collective operators $S_{-}^{\prime }$ and $%
S_{+}^{\prime }$ are the same as those given in Eq.~(22).

One can see that the four Hamiltonians (51-54) obtained here have the same
forms as the Hamiltonians (8), (9), (20), and (21), respectively. Hence, the
NTCP gate can be implemented with charge qubits coupled to a resonator. A
more detailed discussion on this is given in the next subsection.

\begin{center}
\textbf{B. NTCP gates with charge qubits coupled to a resonator}
\end{center}

Following the first method introduced in the previous section (see section
IV.A), we now discuss how to implement the NTCP gate with $(n+1)$ charge
qubits ($1,$ $2,$ $...,$ $n+1$), coupled to a superconducting resonator. To
begin with, it should be mentioned that: (a) for each step of the
operations, the dc gate voltage $V_g^{\mathrm{dc}}$ for each one of qubits ($%
1,$ $2,$ $...,$ $n+1$) is set by $V_g^{\mathrm{dc}}=e/C_g,$ such that $E_z=0$
for each qubit; and (b) the resonator mode frequency $\omega _c$ is \textit{%
\ fixed} during the entire operation. The three-step operations for the gate
realization are illustrated in Fig.~6.

\begin{figure}[tbp]
\includegraphics[bb=95 206 533 752, width=8.6 cm, clip]{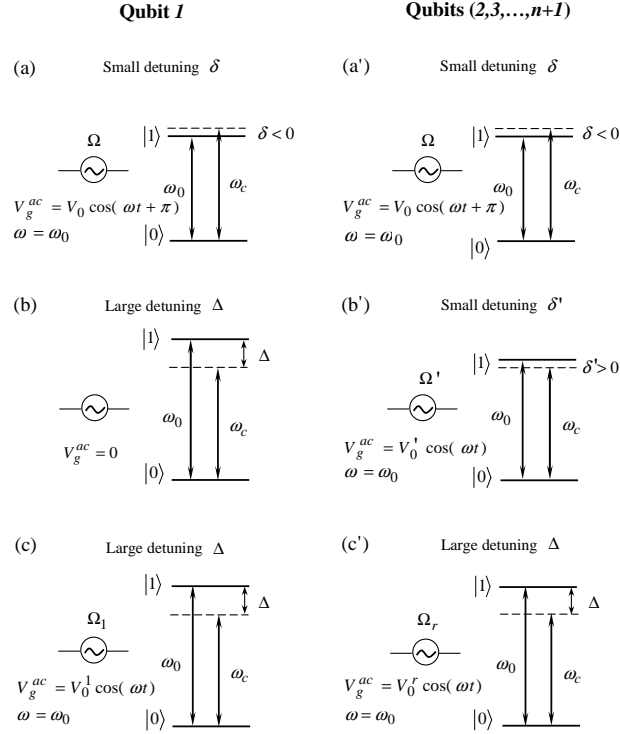} %
\vspace*{-0.18in}
\caption{Change of the qubit transition frequency $\omega _0$ and the ac
gate-voltage frequency $\omega$ during a three-step NTCP gate with charge
qubits coupled to a resonator. The three figures on the left side correspond
to the charge control qubit $1,$ which (from top to bottom) are for the
operations of step (i), step (ii), and step (iii), respectively. The three
figures on the right side correspond to the charge target qubits ($%
2,3,...,n+1$), (from top to bottom) for the operations of step (i), step
(ii), and step (iii), respectively. In each figure, the two horizontal solid
lines represent the qubit levels $\left| 0\right\rangle $ and $\left|
1\right\rangle $; $\omega _c$ is the resonator mode frequency; $\omega _0$
is the qubit transition frequency; $\omega $ is the frequency of the ac gate
voltage $V_g^{\mathrm{ac}}$ (i.e., the pulse); and $\Omega \left( V_0\right)
$, $\Omega ^{\prime }\left( V_0^{\prime }\right) ,$ $\Omega _1\left(
V_0^1\right) $, or $\Omega _r\left( V_0^{\mathrm{r}}\right) $ is the
function of the amplitude $V_0,$ $V_0^{\prime },$ $V_0^1,$ or $V_0^r$ of the
ac gate voltage, which can be adjusted by changing the ac gate-voltage
amplitude; and each circle with a symbol $\sim$ represents an ac gate
voltage. In (b), the ac gate voltage for qubit $1$ is set to zero (i.e., $%
V_g^{\mathrm{ac}}=0$). In addition, $\delta$ and $\delta ^{\prime }$ are the
small detunings of the resonator mode with the $\left| 0\right\rangle
\leftrightarrow \left| 1\right\rangle $ transition, which are given by $%
\delta =\omega _0-\omega _c<0$ and $\delta ^{\prime }=\omega _0-\omega _c>0$%
; while $\Delta=\omega _0-\omega _c$ represents the large detuning of the
resonator mode with the $\left| 0\right\rangle \leftrightarrow \left|
1\right\rangle $ transition. Note that the resonator mode frequency $\omega
_c$ is kept fixed during the entire operation, but the qubit transition
frequency $\omega _0$ is adjusted to achieve a different detuning $\delta,
\delta^{\prime}$, or $\Delta$ for each step.}
\label{fig:6}
\end{figure}

From Fig.~6, it can be seen that there is no need of adjusting the resonator
mode frequency $\omega _c$. Hence, the procedure here for the gate
realization is an extension of the first method introduced above. Note that
the three time-evolution operators $U,$ $U^{\prime }$, and $\widetilde{U}$,
obtained from each step, are the same as those obtained from each step in
subsection IV.A. Therefore, following the same discussion given there, one
can easily see that the NTCP gate can be implemented with $\left( n+1\right)
$ charge qubits (i.e., the control charge qubit $1,$ as well as the $n$
target charge qubits $2,3,...,$ and $n+1$). Namely, after the three-step
process in Fig. 6, a phase flip (i.e., $\left| -\right\rangle \rightarrow
-\left| -\right\rangle $) on the state $\left| -\right\rangle $ of each
target charge qubit is achieved when the control charge qubit $1$ is
initially in the state $\left| -\right\rangle $, but nothing happens to the
states $\left| +\right\rangle $ and $\left| -\right\rangle $ of each target
charge qubit when the control charge qubit $1$ is initially in the state $%
\left| +\right\rangle .$

From the above description, one can see that the spin operator $\widetilde{%
\sigma }_z$ is identical to $\sigma _x.$ In other words, the states $\left|
+\right\rangle $ and $\left| -\right\rangle $ are the eigenstates of both
operators $\widetilde{\sigma }_z$ and $\sigma _x.$ Thus, the NTCP gate
(implemented with charge qubits here) is actually performed with respect to
the eigenstates $\left| +\right\rangle $ and $\left| -\right\rangle $ of the
spin operator $\widetilde{\sigma }_z$ (i.e., the two charge states $n=0$ and
$n=1$ above, for each qubit).

According to the discussion in subsection IV.A, it can be found that to
implement the NTCP gate, the following conditions need to be satisfied: (a) $%
\Omega \gg g,\delta $ and $\Omega ^{\prime }\gg g^{\prime },\delta ^{\prime
};$ (b) $\Omega _1=-ng^2/\delta +\Omega $ and $\Omega _r=-g^2/\delta ;$ (c) $%
-\Omega /\delta =\Omega ^{\prime }/\delta ^{\prime }$ and $-g/\delta
=g^{\prime }/\delta ^{\prime };$ and (d) $4g^2/\delta ^2=2k+1.$ These
conditions can in principle be realized because: (i) The Rabi frequencies $%
\Omega \left( V_0\right) ,$ $\Omega ^{\prime }\left( V_0^{\prime }\right)
,\Omega _1\left( V_0^1\right) ,$ and $\Omega _r\left( V_0^{\mathrm{r}%
}\right) $ are respectively functions of the amplitudes $V_0,V_0^{\prime }$,
$V_0^1,$ and $V_0^{\mathrm{r}}$ of the ac gate voltages, which can be
adjusted by changing the amplitudes of the ac gate voltages; and (ii) The
detunings $\delta $ and $\delta ^{\prime }$ can be adjusted by changing the
qubit transition frequency $\omega _0.$

Note that $n_g^{\mathrm{dc}}=1/2$ (i.e., $E_z=0$) was set for each qubit
during the entire operation. We now give some discussion on the deviation
from the degeneracy point $n_g^{\mathrm{dc}}=1/2$ for each one of the qubits
($1,~2,~...,~n+1$), during each step shown in Fig. (6).

(a) For each one of the qubits ($1,~2,~...,~n+1$) in step (i), we have, from
Eq. (42), that

\begin{equation}
n_g=n_g^{\mathrm{dc}}+n_0^{\mathrm{ac}}\cos \left( \omega t+\varphi \right)
+n_0^{\mathrm{qu}}(a+a^{\dagger }),
\end{equation}
where
\begin{eqnarray*}
n_0^{\mathrm{ac}} &=&C_gV_0/\left( 2e\right) =\hbar \Omega /\left(
4E_c\right) , \\
n_0^{\mathrm{qu}} &=&C_gV_0^{\mathrm{qu}}/\left( 2e\right) =\hbar g/\left(
4E_c\right) .
\end{eqnarray*}
Therefore, the maximal deviation from the degeneracy point for each one of
the qubits ($1,~2,~...,$~$n+1$) in step (i) is
\begin{equation}
\varepsilon _0=\left| n_g^{\mathrm{dc}}+n_0^{\mathrm{ac}}+n_0^{\mathrm{qu}%
}-1/2\right| =\hbar \left( \Omega +g\right) /\left( 4E_c\right) .
\end{equation}

(b) Similarly, one can find that the maximal deviation from the degeneracy
point for qubits ($2,3,...,n+1$) in step (ii) is
\begin{equation}
\varepsilon _1=\hbar \left( \Omega ^{\prime }+g^{\prime }\right) /\left(
4E_c\right) .
\end{equation}
Note that the deviation from the degeneracy point for qubit $1$ is smaller
than $\varepsilon _1$ since $V_g^{\mathrm{ac}}=0$ and thus $n_0^{\mathrm{ac}%
}=0$ for qubit $1.$

(d) For step (iii), we have $n_0^{\mathrm{ac}}=\hbar \Omega _1/\left(
4E_c\right) $ for qubit $1,$ $n_0^{\mathrm{ac}}=\hbar \Omega _r/\left(
4E_c\right) $ for qubits ($2,3,...,n+1$), and $n_0^{\mathrm{qu}}=0.$ Thus,
it is easy to see that the maximal deviation from the degeneracy point for
qubit $1$ is
\begin{equation}
\varepsilon _2=\hbar \Omega _1/\left( 4E_c\right) ,
\end{equation}
and the deviation from the degeneracy point for qubits ($2,3,...,n+1$) is
\begin{equation}
\varepsilon _3=\hbar \Omega _r/\left( 4E_c\right) .
\end{equation}

In the above, we have discussed how to realize the NTCP gate with
superconducting charge qubits coupled to a resonator. Note that the proposal
here can implement multi-qubit gates, while previous proposals
(e.g.,[49-51]) using superconducting qubits are limited to two-qubit gates.

\begin{center}
\textbf{C. Possible experimental implementation}
\end{center}

In this section we discuss some issues which are relevant for future
experimental implementation of our proposal. For the method to work: (a) The
conditions for the Rabi frequencies $\Omega ,$ $\Omega ^{\prime },$ $\Omega
_1,$ and $\Omega _r,$ which were discussed above, need to be met; (b) The
total operation time
\begin{equation}
t_{\mathrm{op}}=2\tau +\tau ^{\prime }=4\pi /\left| \delta \right| +2\pi
/\delta ^{\prime }
\end{equation}
should be much shorter than the energy relaxation time $T_1$ and the
dephasing time $T_2$ of the qubit and the lifetime of the resonator mode $%
\kappa ^{-1}=Q/\omega _c,$ where $Q$ is the (loaded) quality factor of the
resonator; (c) The deviations $\varepsilon _0,$ $\varepsilon _1,$ $%
\varepsilon _2,$ and $\varepsilon _3$ from the degeneracy point need to be
small numbers to have the qubits working near the degeneracy point, such
that the qubits are less affected by the low-frequency charge noises
[52,53]; and (d) The direct coupling between SQUIDs needs to be negligible,
since this interaction is not intended. It is noted that the direct
interaction between SQUIDs can be made negligibly small as long as $D\gg d$
(where $D$ is the distance between the two nearest SQUIDs and $d$ is the
linear dimension of each SQUID).

\begin{figure}[tbp]
\includegraphics[bb=16 280 536 577, width=8.6 cm, clip]{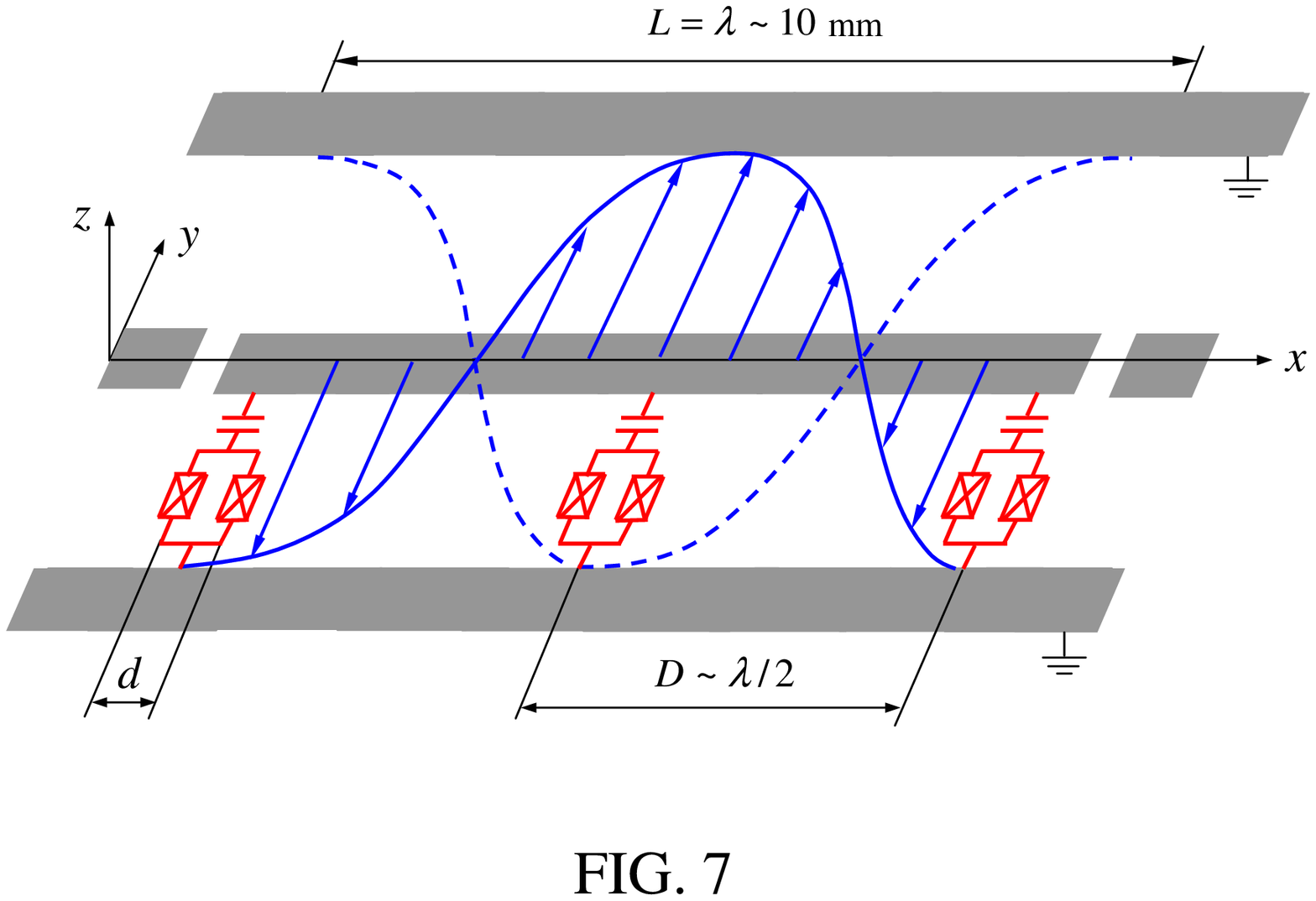} %
\vspace*{-0.18in}
\caption{(Color online) Proposed setup for three qubits (red squares) and a
(grey) standing-wave quasi-one dimensional coplanar waveguide cavity (not
drawn to scale). Each qubit is placed at an antinode of the electric field.
The two blue curves represent the standing-wave electric field, along the $y$%
-direction. $V_g$ is the gate voltage, $D$ is the distance between any two
nearest SQUID loops, and $d$ is the linear dimension of each SQUID loop.}
\label{fig:7}
\end{figure}

\begin{table}[tbp]
\includegraphics[bb=73 276 389 478, width=8.6 cm, clip]{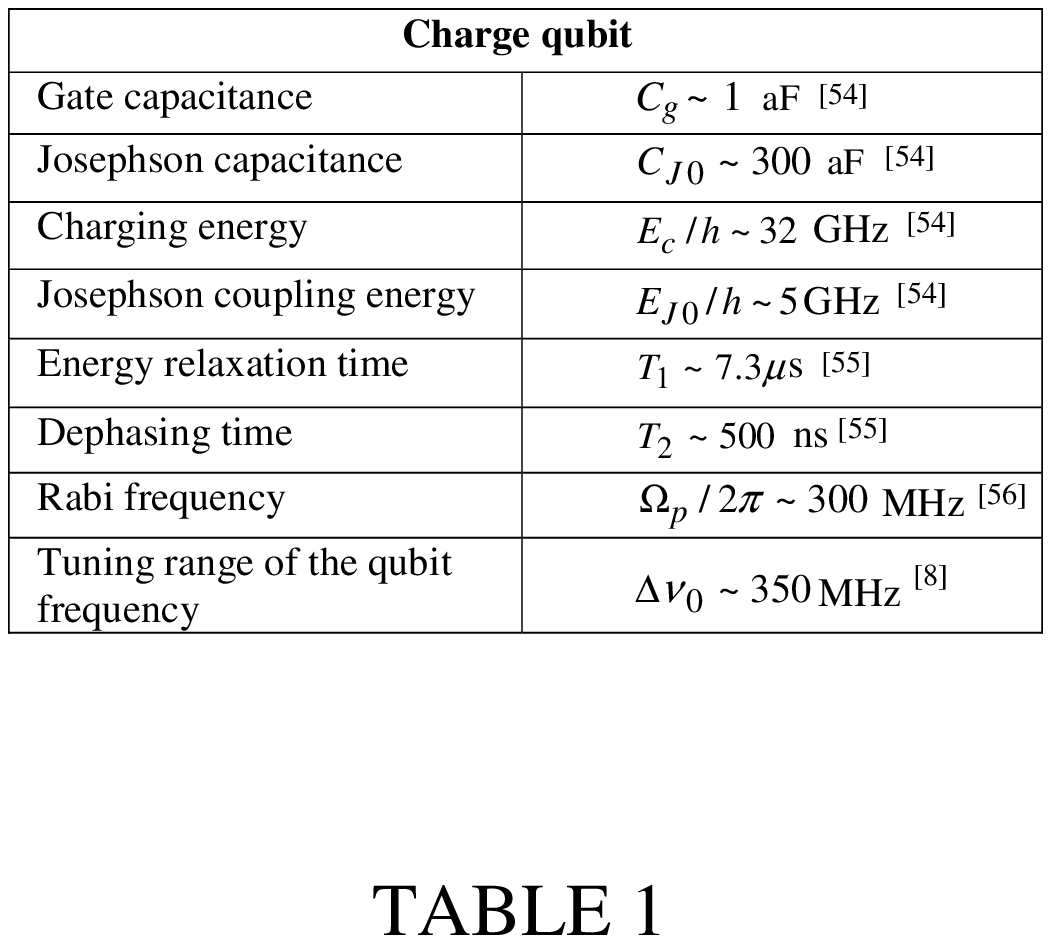} %
\vspace*{-0.18in}
\caption{Possible experimental parameters of a charge qubit [8,54-56]. Here,
$\Omega_p$ is the typical Rabi frequency available in experiments, and $%
\Delta \nu _0$ is the experimentally tuning range of the qubit frequency.}
\label{table:1}
\end{table}

For the sake of definitiveness, let us consider the experimental feasibility
of implementing a two-target-qubit controlled phase gate using
superconducting charge qubits with parameters listed in Table I [8,54-56].
Note that in a recent experiment, coupling three superconducting qubits with
a transmission line resonator has been demonstrated [57]. For a
superconducting one dimensional standing-wave CPW (coplanar waveguide)
transmission line resonator and each qubit placed at an antinode of the
resonator mode (Fig.~7), the amplitude of the quantum part of the gate
voltage is given by [48]
\begin{equation}
V_0^{\mathrm{qu}}=\left( \hbar \omega _c\right) ^{1/2}\left( Lc_0\right)
^{-1/2},
\end{equation}
where $L$ is the length of the resonator and $c_0$ is the capacitance per
unit length of the resonator. Therefore, the coupling constant $g$ is given
by
\begin{equation}
g=2E_cC_g\left( \hbar e\right) ^{-1}\left( \hbar \omega _c\right)
^{1/2}\left( Lc_0\right) ^{-1/2},
\end{equation}
showing that $g$ does not depend on the detuning $\delta $. Therefore, we
have $g=g^{\prime }$, for which the above condition $-\Omega /\delta =\Omega
^{\prime }/\delta ^{\prime }$ and $-g/\delta =g^{\prime }/\delta ^{\prime }$
simply turns to the $\Omega =\Omega ^{\prime }$ and $-\delta =\delta
^{\prime }.$ For superconducting charge qubits with parameters given in
Table I, and a resonator with the parameters listed in Table II [36,57-59],
a simple calculation gives $g/2\pi \sim 22$ MHz, which is available in
experiments (see, e.g., [58]). With a choice of $-\delta =\delta ^{\prime
}=2g$ [corresponding to $k=0$ in Eq. (36)], the total operation time $t_{%
\mathrm{op}}$ would be $\sim 68$ ns, which is much shorter than the
dephasing time $T_2$ and $\kappa ^{-1}\sim 794$ ns for a resonator with $%
Q=10^5$. Note that a superconducting CPW resonator with a quality factor of $%
Q>10^6$ has been experimentally demonstrated [59].

\begin{table}[tbp]
\includegraphics[bb=74 444 430 612, width=8.6 cm, clip]{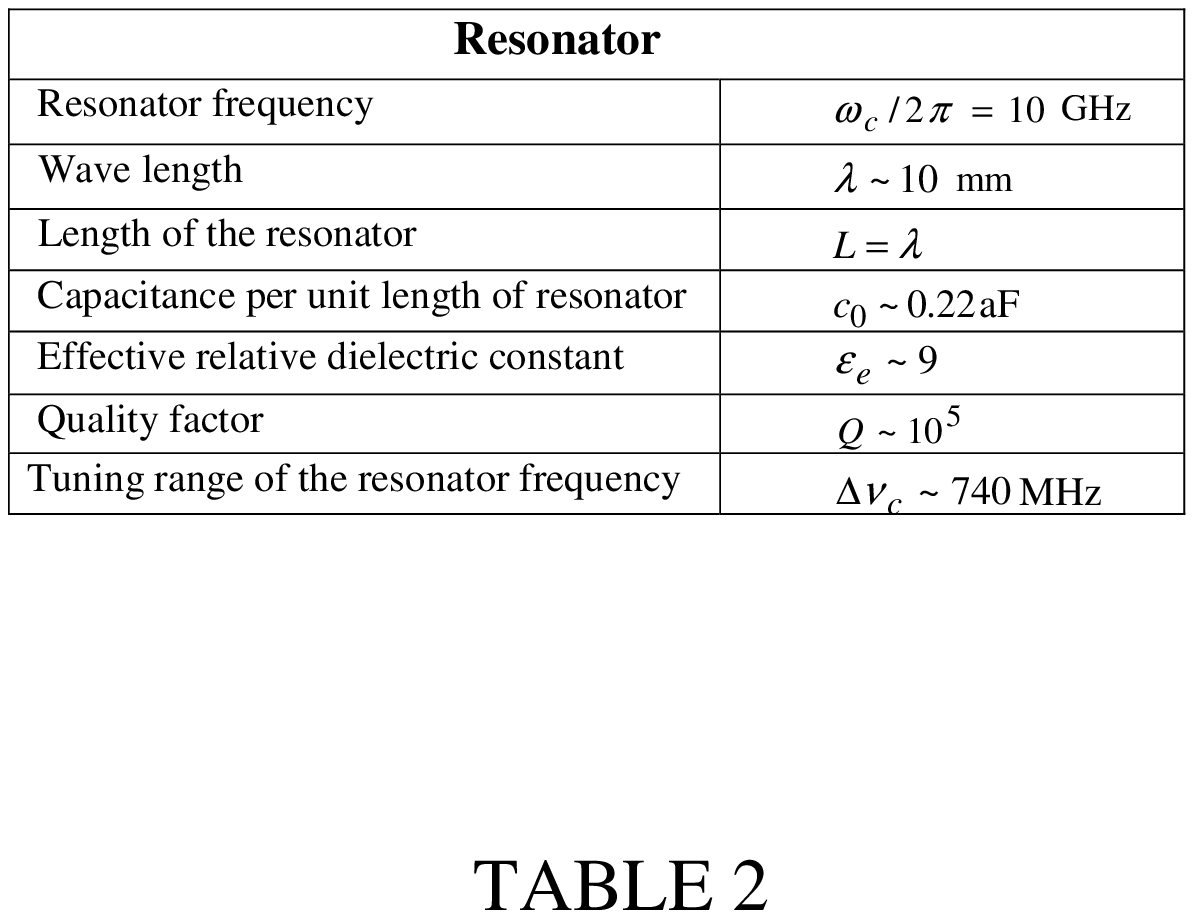} %
\vspace*{-0.18in}
\caption{Possible experimental parameters of a resonator [36,57-59]. Here, $%
\Delta \nu _c$ is the experimentally tuning range of the resonator
frequency. }
\label{table:2}
\end{table}

Based on $\Omega =$ $\Omega ^{\prime }$, $\Omega _1=4\lambda n+\Omega ,$ and
$\Omega _r=4\lambda $ ($n=2$ for a two-target-qubit gate), we have $\Omega
^{\prime }/\left( 2\pi \right) \sim 330$ MHz, $\Omega _1/\left( 2\pi \right)
\sim 352$ MHz, and $\Omega _r/\left( 2\pi \right) \sim 11$ MHz for a choice
of $\Omega \sim 15g,$ i.e., $\Omega /\left( 2\pi \right) \sim 330$ MHz. For
the Rabi frequencies given here, we obtain $\varepsilon _0=$ $\varepsilon
_1\sim 2.75\times 10^{-3},$ $\varepsilon _2\sim 2.75\times 10^{-3},$ and $%
\varepsilon _3\sim 8.59\times 10^{-5}$ for a qubit-cavity system with the
above parameters. Therefore, the conditions for the qubits to work near the
degeneracy point are well satisfied.

Finally, for a resonator with $\omega _c/\left( 2\pi \right) =10$ GHz, the
wavelength of the resonator mode is $\lambda \sim 10$ mm. For the charge
qubits placed in a resonator as shown in Fig. 7, the distance between any
two nearest SQUIDs is $D\sim \lambda /2\sim 5$ mm. Hence, the ratio $D/d$
would be $\sim 250$ for $d=20\mu $m. Note that the dipole field generated by
the current in each SQUID ring at a distance $r\gg d$ decreases as $r^{-3}$.
Thus, the condition of negligible direct coupling between SQUIDs is well
satisfied.

Note that because of $\omega _0=4E_{J0}\cos \left( \pi \Phi /\Phi _0\right)
/\hbar ,$ it can be found that for charge qubits with the parameters given
in Table I, the transition frequency $\nu _0=\omega _0/2\pi $ of each qubit
varies from $\nu _0=0$ GHz for $\Phi /\Phi _0=1/2$ to $\nu _0=20$ GHz for $%
\Phi =0$. Therefore, the choice of the resonator frequency above is
reasonable. For the choice of $-\delta =\delta ^{\prime }=2g$ above, the
qubit transition frequency $\nu _0$ would be $\sim 9.956$ GHz for the
detuning $\delta =-2g$ while $\sim 10.044$ GHz for the detuning $\delta
^{\prime }=2g.$

The above analysis shows that the realization of a two-target-qubit
controlled phase gate is possible using superconducting charge qubits and a
resonator. We remark that a quantum-controlled phase gate with a larger
number of target qubits can in principle be obtained by increasing the
length of the resonator since the total operation time $t_{\mathrm{op}}$ is
independent of the number of target qubits $n.$

\begin{figure}[tbp]
\includegraphics[bb=8 233 505 685, width=8.6 cm, clip]{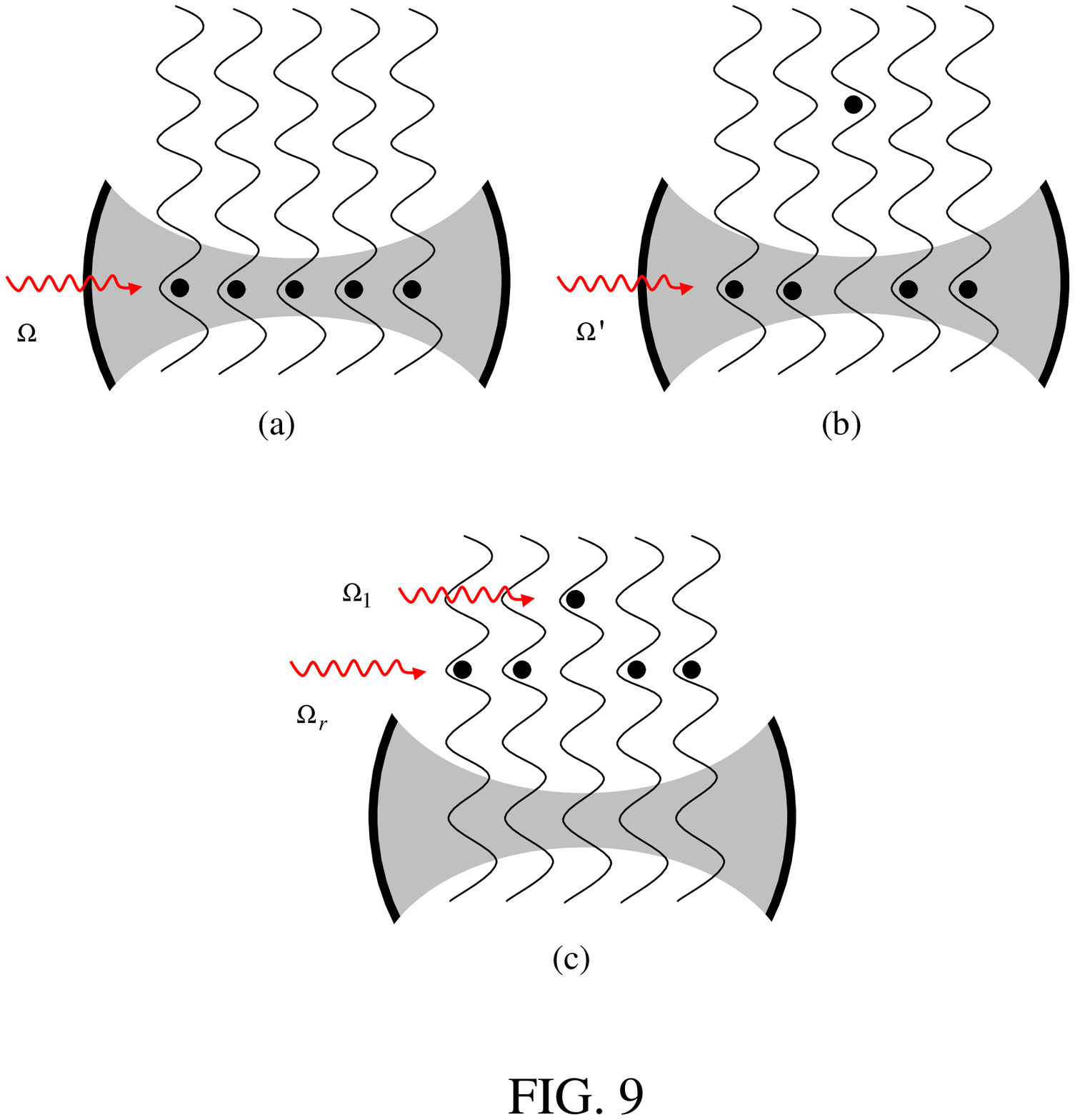} %
\vspace*{-0.18in}
\caption{(Color online) Proposed set-up for an $n$-target-qubit control
phase (NTCP) gate with ($n+1$) identical neutral atoms and a cavity. For
simplicity, only five atoms are drawn here. Each atom can be either loaded
into the cavity or moved out of the cavity by one-dimensional translating
optical lattices [20,45,46]. The atom in the middle represents atom 1 (the
control qubit), while the remaining atoms play the role of target qubits.}
\label{fig:8}
\end{figure}

\begin{center}
\textbf{VI. NTCP GATE WITH ATOMS USING ONE CAVITY}
\end{center}

Consider $\left( n+1\right) $ identical two-level atoms ($1,2,...,n+1$). The
two levels of each atom are labelled by $\left| 0\right\rangle $ and $\left|
1\right\rangle .$ The transition frequency of each atom is denoted as $%
\omega _0.$ Each atom is trapped in the periodic potential of a
one-dimensional optical lattice and can be loaded into or moved out of the
cavity by translating the optical lattice [20,45,46]. The NTCP gate can be
realized using a procedure illustrated in Fig. 8. The operations shown in
Fig. 8 are as follows:

$\bullet $ Figure 8(a): Move atoms ($1,2,...,n+1$) into the cavity and then
apply a classical pulse (with an initial phase $\varphi =\pi $ and frequency
$\omega =\omega _0$) to the atoms. The cavity mode is coupled to the $\left|
0\right\rangle \rightarrow $ $\left| 1\right\rangle $ transition of each
atom, with a detuning $\delta =\omega _0-\omega _c<0$, which can be achieved
by prior adjustment of the cavity mode frequency [35]. The time-evolution
operator for this operation is the $U$ in Eq.~(19) for an interaction time $%
\tau =-2\pi /\delta .$

$\bullet $ Figure 8(b): Move atoms ($1,2,...,n+1$) out of the cavity and
then adjust the cavity mode frequency. After adjusting the cavity mode
frequency, move atoms ($2,3,...,n+1$) back into the cavity and then apply a
classical pulse (with $\varphi =0$ and $\omega =\omega _0$) to them. The
cavity mode frequency is adjusted such that the cavity mode is coupled to
the $\left| 0\right\rangle \rightarrow $ $\left| 1\right\rangle $ transition
of atoms ($2,3,...,n+1$), with a detuning $\delta ^{\prime }=\omega
_0-\omega _c>0$. The time-evolution operator for this operation is the $%
U^{\prime }$ in Eq.~(26) for an interaction time $\tau ^{\prime }=2\pi
/\delta ^{\prime }.$

$\bullet $ Figure 8(c): Move atoms ($2,3,...,n+1$) out of the cavity. Apply
a classical pulse (with $\varphi =0$ and $\omega =\omega _0$) to the control
atom $1$ and a classical pulse (with the same initial phase and frequency)
to the target atoms ($2,3,...,n+1$). The Rabi frequency for the pulse
applied to atom $1$ is $\Omega _1$, while the Rabi frequency for the pulses
applied to atoms ($2,3,...,n+1$) is $\Omega _r.$ This operation is described
by the operator $\widetilde{U}$ in Eq.~(28) for an interaction time $\tau $.

The three time-evolution operators $U,$ $U^{\prime }$, and $\widetilde{U}$
here are the same as those obtained from each step in subsection IV.A.
Therefore, as long as the conditions given in subsection IV.A are met, an
NTCP gate described by Eq.~(39) is implemented with the $\left( n+1\right) $
atoms. Namely, after the operations shown in Fig. 8, $n$ two-qubit
controlled phase gates, each described by the unitary transformation in
Eq.~(1), are performed on the atom pairs ($1,2$), ($1,3$),..., and ($1,n+1$%
), respectively. Note that each pair contains the same control qubit (atom $%
1 $) but a different target qubit (atom $2,3,...,$ or $n+1$).

As shown above, the present scheme for implementing the NTCP gate with atoms
has the following features:

(a) No adjustment of the level spacings for each atom is required during the
entire operation;

(b) Only one cavity is required;

(c) The cavity mode can be initially in an arbitrary state; and

(d) The operation time does not depend on the number of atoms.

For our scheme to work, the total operation time $t_{\mathrm{op}}=\tau +\tau
^{\prime }+\tau _{\mathrm{a}}+4\tau _{\mathrm{m}}$ should be much smaller
than the cavity decay time $\kappa ^{-1}$, so that the cavity dissipation is
negligible. Here, $\tau _{\mathrm{a}}$ is the typical time required for
adjusting the cavity mode frequency during step (ii) above, and $\tau _{%
\mathrm{m}}$ is the typical time required for moving atoms into or out of
the cavity. In addition, the $t_{\mathrm{op}}$ needs to be much smaller than
the energy relaxation time of level $\left| 1\right\rangle $, such that the
decoherence induced due to the spontaneous decay of the level $\left|
1\right\rangle $ is negligible. In principle, these conditions can be
satisfied by choosing a cavity with a high quality factor $Q$ and atoms with
a sufficiently long energy relaxation time.

To investigate the experimental feasibility of this proposal, let us
consider Rydberg atoms with principal quantum numbers 50 and 51
(respectively corresponding to the levels $\left| 0\right\rangle $ and $%
\left| 1\right\rangle $). The $\left| 0\right\rangle \leftrightarrow \left|
1\right\rangle $ transition frequency is $\omega _0/2\pi \sim 51.1$ GHz, the
energy relaxation time of the level $\left| 1\right\rangle $ is $T_r\sim
3\times 10^{-2}$ s, and the coupling constant is $g=2\pi \times 50$ KHz
[35,60]. Now set $-\delta =\delta ^{\prime }=2g$, and assume $g\sim
g^{\prime }$. With the choice of these parameters, the time needed for the
entire operation is $t_{\mathrm{op}}\sim $ 65 $\mu $s for $\tau _{\mathrm{a}%
}=\tau _{\mathrm{m}}\sim 1$ $\mu $s, which is much shorter than $T_r$. The
cavity mode frequency is $\omega _c/2\pi \sim 51.2$ GHz for the negative
qubit detuning $\delta =-2g$ while $\sim 51$ GHz for the positive qubit
detuning $\delta ^{\prime }=2g.$ To estimate the lifetime of the cavity
photon, we here consider the conservative case of a larger cavity frequency $%
\omega _c/2\pi \sim 51.2$ GHz, for which the lifetime of the cavity photon
is $T_c=Q/\omega _c\sim 622$ $\mu $s for a cavity with $Q=2\times 10^8$,
which is much larger than the total operation time $t_{\mathrm{op}}.$ Note
that cavities with a high $Q\sim 10^{10}$ have been demonstrated in
experiments [61]. Thus, the present proposal might be realizable using
current cavity QED setups.

\begin{center}
\textbf{VII. CONCLUSIONS}
\end{center}

We have presented two different methods for the proposed NTCP gate
implementation. The two methods are quite general, which can be applied to
physical systems such as trapped atoms, quantum dots, and superconducting
qubits. For the two methods, we have provided guidelines on how to protect
multi-level qubits from leaking out of the computational subspace.

Using a concrete example, we have shown how to apply the first method to
implement the NTCP gate with superconducting qubits coupled to a resonator.
In addition, we have discussed the experimental feasibility of performing a
two-target-qubit controlled phase gate with superconducting qubits coupled
to a one-dimensional transmission line resonator. Our analysis shows that
the realization of this gate is possible within current technologies. How
well this gate would work in light of experimental errors should be further
investigated elsewhere for each particular set-up or implementation. This is
beyond the scope of this theoretical work.

We have shown how to extend the second method to implement the NTCP gate
with trapped atomic qubits by using one cavity. Interestingly, as shown
above, there is no need to adjust the level spacings of atoms during the
entire operation, and decoupling of atoms with the cavity can be easily
achieved by just moving atoms out of the cavity.

In summary, we have presented a general proposal to implement a NTCP gate
with qubits in a cavity or coupled to a resonator. As shown above, the
present proposal has the following features: (i) The $n$ two-qubit CP gates
involved in the NTCP gate can be simultaneously performed; (ii) The
operation time required for the gate implementation is independent of the
number of the target qubits, thus it does not increase with the number of
qubits; (iii) The gate is insensitive to the initial state of the cavity
mode, therefore no preparation for a specified initial state of the cavity
mode is needed; (iv) No measurement on the qubits or the cavity mode is
needed and thus the operation is simplified, (v) The gate realization
requires only three steps of operations.

\begin{center}
\textbf{ACKNOWLEDGMENTS}
\end{center}

FN and CPY acknowledge partial support from the National Security Agency
(NSA), Laboratory for Physical Sciences (LPS), (U.S.) Army Research Office
(USARO), National Science Foundation (NSF) under Grant No. 0726909, and
JSPS-RFBR under Contract No. 06-02-91200. YXL is supported by the National
Natural Science Foundation of China under Grant Numbers 10975080 and
60836001.

\begin{center}
\textbf{APPENDIX: HOW TO PROTECT QUBITS FROM LEAKING TO HIGHER ENERGY LEVELS}
\end{center}

Let us now discuss how to protect qubits from leaking out of the
computational subspace in the presence of more qubit levels. Generally
speaking, we need to consider two situations, which are here denoted as
cases $L $ and $S$. For case $L$, the level spacing between the level $%
\left| 1\right\rangle $ and the level $\left| 2\right\rangle $ (the first
level above $\left| 1\right\rangle $) is \textit{larger} than the level
spacing between the levels $\left| 0\right\rangle $ and $\left|
1\right\rangle .$ Namely, for case $L$, $\left( E_2-E_1\right) >\left(
E_1-E_0\right) ,$ where $E_0,$ $E_1,$ and $E_2$ are the energy eigenvalues
of the levels $\left| 0\right\rangle ,\left| 1\right\rangle ,$ and $\left|
2\right\rangle .$ For case $S$, the level spacing between the levels $\left|
1\right\rangle $ and $\left| 2\right\rangle $ is \textit{smaller} than that
between the levels $\left| 0\right\rangle $ and $\left| 1\right\rangle .$
Namely, $\left( E_2-E_1\right) <\left( E_1-E_0\right) .$ For solid-state
qubits, case $L$ exists for superconducting charge qubits and flux qubits
[31]; while case $S$ can be applied to superconducting phase qubits [33].
For atomic qubits, the level structures for both cases $L$ and $S$ are
available.

Here and below, we define $\Delta _2=\left( E_2-E_1\right) /\hbar -\omega _c$%
, $\Delta _3=\left( E_3-E_1\right) /\hbar -\omega _c$, and $\Delta _a=\left(
E_a-E_1\right) /\hbar -\omega _c$, as the large detuning of the cavity mode
with the $\left| 1\right\rangle \leftrightarrow \left| 2\right\rangle $
transition, the large detuning of the cavity mode with the $\left|
1\right\rangle \leftrightarrow \left| 3\right\rangle $ transition, and the
large detuning of the cavity mode with the $\left| 1\right\rangle
\leftrightarrow \left| a\right\rangle $ transition, respectively. $E_a$ is
the energy eigenvalue of level $\left| a\right\rangle $.

\begin{center}
\textbf{A. Case $L$: ~$\left( E_2-E_1\right) >\left( E_1-E_0\right) $}
\end{center}

For case $L,$ let us consider Fig. 9(a), where the dashed line falls within
the range between the levels $\left| 1\right\rangle $ and $\left|
2\right\rangle $ when the cavity mode is slightly detuned from the $\left|
0\right\rangle \leftrightarrow \left| 1\right\rangle $ transition.
Therefore, a large detuning $\Delta _2$ is needed to avoid the transition
from the level $\left| 1\right\rangle $ to the level $\left| 2\right\rangle $%
. Note also that as long as this large detuning is met, no transition from
level $\left| 1\right\rangle $ to the higher energy level above the level $%
\left| 2\right\rangle $ happens. To explain this, let us consider an
arbitrary level $\left| a\right\rangle $ above level $\left| 2\right\rangle $
[Fig.~9(a)]. Since the detuning $\Delta _a$ is larger than the detuning $%
\Delta _2$, then the large detuning regime for $\Delta _a$ is automatically
satisfied when $\Delta _2$ is large. As a result, the transition from the
level $\left| 1\right\rangle $ to the level $\left| a\right\rangle $ will
not be induced by the cavity mode. In addition, the excitation of the level $%
\left| a\right\rangle ,$ induced due to its coupling to the level $\left|
2\right\rangle $ via the cavity mode, does not occur when the level $\left|
2\right\rangle $ is unpopulated. Hence, as long as the large detuning $%
\Delta _2$ is satisfied, no level above $\left| 1\right\rangle $ will be
populated and therefore the leak out of the computational subspace is
suppressed.

\begin{figure}[tbp]
\includegraphics[bb=159 270 486 569, width=8.6 cm, clip]{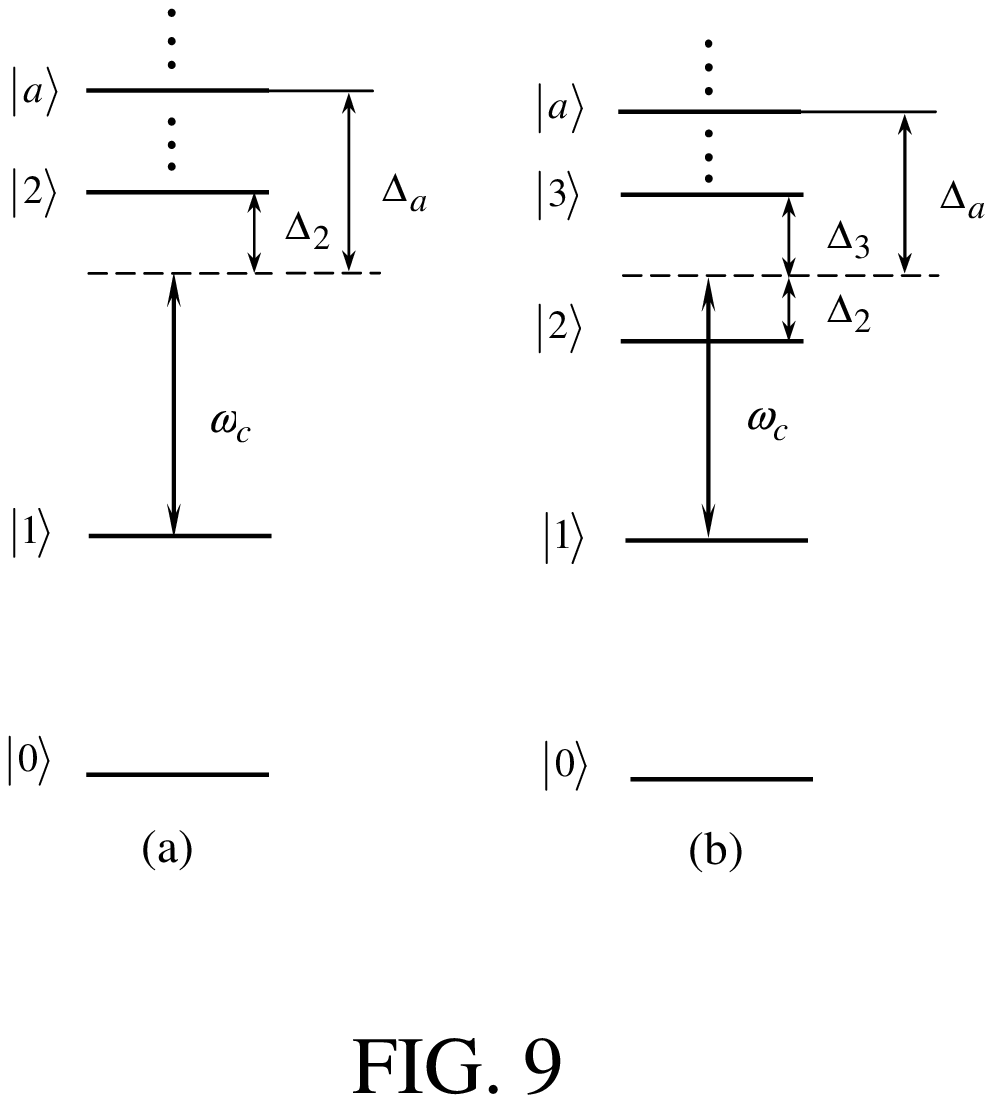} %
\vspace*{-0.18in}
\caption{Large detuning of the cavity mode with the transition between
energy levels. Figure (a) corresponds to the case $L$, while figure (b)
corresponds to case $S.$ The dots above level $\left| 2\right\rangle $ in
(a) and level $\left| 3\right\rangle $ in (b) represent other energy levels.}
\label{fig:9}
\end{figure}

\begin{center}
\textbf{B. Case $S$: ~$\left( E_2-E_1\right) <\left( E_1-E_0\right) $}
\end{center}

For case $S,$ let us consider Fig. 9(b), where the dashed line falls within
the range between the levels $\left| 2\right\rangle $ and $\left|
3\right\rangle $, when the cavity mode is slightly detuned from the $\left|
0\right\rangle \leftrightarrow \left| 1\right\rangle $ transition. This can
be achieved, e.g., for superconducting qubits by appropriately choosing the
device parameters and/or adjusting the external parameters to control the
level structures. From Fig.~9(b), it can be seen that a large detuning $%
\Delta _2$ and a large detuning $\Delta _3$ are needed to avoid the
transition from the level $\left| 1\right\rangle $ to the level $\left|
2\right\rangle $ or the level $\left| 3\right\rangle $. Note also that as
long as these two large detunings are met, no transition from the level $%
\left| 1\right\rangle $ to the higher energy level above the level $\left|
3\right\rangle $ occurs. To understand this, let us consider an arbitrary
level $\left| a\right\rangle $ above $\left| 3\right\rangle $ [Fig. 9(b)].
Note that the detuning $\Delta _a$ is larger than the detuning $\Delta _3$.
Therefore, the large detuning regime for $\Delta _a$ is automatically
satisfied when the cavity mode is largely detuned from the $\left|
1\right\rangle $ $\leftrightarrow \left| 3\right\rangle $ transition. As a
result, the transition from level $\left| 1\right\rangle $ to level $\left|
a\right\rangle $ will not be induced by the cavity mode. In addition, the
excitation of the level $\left| a\right\rangle $, caused due to its coupling
to the level $\left| 2\right\rangle $ or $\left| 3\right\rangle $ via the
cavity mode, will not happen when levels $\left| 2\right\rangle $ and $%
\left| 3\right\rangle $ are unpopulated. Hence, as long as the large
detunings $\Delta _2$ and $\Delta _3$ are met, the levels above $\left|
1\right\rangle $ will not be excited by the cavity mode.

We now give a quantitative analysis on the effect of the finite detuning of
the qubit frequencies with the cavity mode. For simplicity, we consider the
case where the cavity mode is in a single photon state and the qubit is in
the state $\left| 1\right\rangle .$ It is estimated that the occupation
probability $p_2$ for level $\left| 2\right\rangle $ and the occupation
probability $p_3$ for level $\left| 3\right\rangle $ (induced by the photon)
would be on the order of $4g_{12}^2/\left( 4g_{12}^2+\Delta _2^2\right) $
and $4g_{13}^2/\left( 4g_{13}^2+\Delta _3^2\right) ,$ respectively. Here, $%
g_{12}$ is the coupling constant between the cavity mode and the $\left|
1\right\rangle \leftrightarrow \left| 2\right\rangle $ transition, while $%
g_{13}$ is the coupling constant between the cavity mode and the $\left|
1\right\rangle \leftrightarrow \left| 3\right\rangle $ transition. With a
choice of $\Delta _2=10g_{12}$ and $\Delta _3=10g_{13},$ we have $p_2,$ $%
p_3\sim 0.04,$ which can be further reduced by increasing the detuning $%
\Delta _2$ or $\Delta _3.$ Therefore, the population probability of the
level $\left| 2\right\rangle $ or $\left| 3\right\rangle $ of the qubit can
be made negligible by choosing the detuning appropriately.

\end{document}